\documentclass[12pt]{article}
\usepackage{amsmath,amssymb,epsfig}
\makeatletter \@addtoreset{equation}{section} \makeatother
\renewcommand{\theequation}{\thesection.\arabic{equation}}
\newcommand{\ba}{\begin{array}}
\newcommand{\ea}{\end{array}}
\newcommand{\beq}{\begin{equation}}
\newcommand{\eeq}{\end{equation}}
\newcommand{\bea}{\begin{eqnarray}}
\newcommand{\eea}{\end{eqnarray}}

\def\bce{\begin{center}}
\def\ece{\end{center}}

\def\nonu{\nonumber}

\def\be{\beta}

\newcommand{\tr}{\mbox{Tr}}

\def\eps6{{\displaystyle \mathop{\epsilon}^{6}}{}}

\def\nab6{{\displaystyle \mathop{\nabla}^{6}}{}}

\def\0{{\sst{(0)}}}
\def\1{{\sst{(1)}}}
\def\2{{\sst{(2)}}}
\def\3{{\sst{(3)}}}
\def\4{{\sst{(4)}}}
\def\5{{\sst{(5)}}}
\def\6{{\sst{(6)}}}
\def\7{{\sst{(7)}}}
\def\8{{\sst{(8)}}}

\def\ba{\begin{array}}
\def\ea{\end{array}}
\def\beq{\begin{equation}}
\def\eeq{\end{equation}}
\def\be{\begin{equation}}
\def\ee{\end{equation}}

\def\tr{\mathop{\rm tr}}

\def\eps{\epsilon}

\def\ba{\begin{array}}
\def\ea{\end{array}}
\def\beq{\begin{equation}}
\def\eeq{\end{equation}}
\def\be{\begin{equation}}
\def\ee{\end{equation}}

\def\tr{\mathop{\rm tr}}

\def\eps{\epsilon}

\newcommand{\bean}{\begin{eqnarray*}}
\newcommand{\eean}{\end{eqnarray*}}

\begin{document}
\thispagestyle{empty} \addtocounter{page}{-1}
   \begin{flushright}
\end{flushright}

\vspace*{1.3cm}

\centerline{ \Large \bf  Meta-Stable Brane Configurations  }
\vspace{.3cm} 
\centerline{ \Large \bf  by Dualizing the Two Gauge Groups  } 
\vspace*{1.5cm}
\centerline{{\bf Changhyun Ahn} 
} 
\vspace*{1.0cm} 
\centerline{\it 
Department of Physics, Kyungpook National University, Taegu
702-701, Korea} 
\vspace*{0.8cm} 
\centerline{\tt ahn@knu.ac.kr} 
\vskip2cm

\centerline{\bf Abstract}
\vspace*{0.5cm}
 
We consider 
the ${\cal N}=1$ supersymmetric gauge theories with product gauge groups.
The two kinds of D6-branes in the electric theory are both displaced and rotated 
respectively where these deformations 
are interpreted as the mass terms and quartic terms
for the two kinds of flavors. Then we apply the Seiberg dual to the whole
gauge group factors by moving the branes 
and obtain the corresponding dual gauge theories. 
By analyzing the magnetic superpotentials consisting of an interaction
term between a magnetic meson field and dual matters as well as the above
deformations for each gauge group, we present the type IIA
nonsupersymmetric meta-stable brane configurations.

\baselineskip=18pt
\newpage
\renewcommand{\theequation}
{\arabic{section}\mbox{.}\arabic{equation}}

\section{Introduction}

The dynamical supersymmetry breaking in meta-stable vacua \cite{ISS,IS} 
arises
in the ${\cal N}=1$ supersymmetric gauge theory with massive fundamental 
quarks.  
The additional  quartic
term for the quarks in the 
electric superpotential \cite{GK0710,GK0710-1} also leads to  
the nonsupersymmetric meta-stable ground states in its magnetic theory
when 
the gravitational attraction of NS5-brane \cite{GK} is considered.
In this construction, 
taking the Seiberg dual, magnetic theory,  from an electric theory 
is a crucial step to find out new meta-stable supersymmetry breaking vacua. 

So far, the Seiberg dual one takes in the context of
nonsupersymmetric meta-stable ground states is 
only for a single gauge group
from a single or multiple electric gauge group.   
Although the electric theory has many gauge group factors, 
the magnetic dual only for one single gauge group is considered.
On the other hand, 
it is known, in the construction of 
supersymmetric ground states or its type IIA brane
configurations \cite{GK98},  
that  a number of gauge theory duals(magnetic theory) 
involving
product gauge groups can be interpreted in terms of branes of type IIA
string theory. 
Then it is natural to ask what happens for the dynamical supersymmetry
breaking in meta-stable vacua, in the ${\cal N}=1$ supersymmetric
product gauge theories
which have mass terms \cite{ISS,OO1,FGU,BGHSS} and quartic terms
\cite{GK0710,GK0710-1} 
for the flavors 
in the electric superpotential, if one dualizes the whole two gauge
groups, not a single gauge group.   

One simplest example can be realized by three NS-branes, D4-branes and
D6-branes \cite{BH}. Now the second example can be obtained 
by  adding  orientifold 4-plane to this brane
configuration and describes different gauge group and matter 
contents \cite{Tatar,Ahn97}. 
Or if one adds orientifold 6-plane to the simplest brane configuration, 
one possible third example is realized by four 
NS-branes, D4-branes and D6-branes \cite{LO}.
All of these examples possess their Seiberg duals either in the gauge
theory side \cite{ILS,BIWW,LO} 
or string theory side for the supersymmetric ground states
10 years ago.

In this paper, one reexamines these supersymmetric brane configurations
and extracts the possible brane motions, during the dual process, for 
new meta-stable brane configurations, along the lines of 
\cite{Ahn07-11,Ahn08-1,Ahn08-1two,Ahn08-2,Ahn08-3}. 
The geometrical positions of the
branes and the creation of D4-branes when the NS5-brane and D6-brane
are intersecting each other with an angle, play the important role for
removing the unwanted gauge singlets and selecting 
the wanted gauge singlet which is originated from the quadratic 
term and mass term of flavors in an electric theory.  

In section 2, we review the type IIA brane configuration corresponding
to the ${\cal N}=1$ $SU(N_c) \times SU(N_c')$ gauge theory 
with fundamentals and bifundamentals and deform this theory 
by adding both the mass terms
and the quartic terms for the fundamentals. 
Then we describe the dual 
${\cal N}=1$ $SU(\widetilde{N}_c) \times SU(\widetilde{N}_c')$ gauge 
theory with corresponding dual
matter as well as a gauge singlet. 
We discuss the nonsupersymmetric meta-stable
minimum  and present 
the corresponding 
intersecting brane configuration of type IIA string
theory.

In section 3,
we review the type IIA brane configuration corresponding
to the ${\cal N}=1$ $SO(2N_c) \times Sp(N_c')$ gauge theory 
with vectors, fundamentals, and bifundamentals and deform this theory 
by adding both the mass terms
and the quartic terms for the vectors and fundamentals. 
Then we describe the dual 
${\cal N}=1$ $SO(2\widetilde{N}_c) \times Sp(\widetilde{N}_c')$ gauge 
theory with corresponding dual
matter as well as a gauge singlet. 
We describe the nonsupersymmetric meta-stable
minimum  and 
the corresponding 
intersecting brane configuration of type IIA string
theory.

In section 4,
we review the type IIA brane configuration corresponding
to the ${\cal N}=1$ $SU(N_c) \times SO(N_c')$ gauge theory 
with fundamentals, vectors, and bifundamentals and deform this theory 
by adding both the mass terms
and the quartic terms for the fundamentals and vectors. 
Then we describe the dual 
${\cal N}=1$ $SU(\widetilde{N}_c) \times SO(\widetilde{N}_c')$ gauge 
theory with corresponding dual
matter as well as a gauge singlet. 
We study the nonsupersymmetric meta-stable
minimum  and 
the corresponding 
intersecting brane configuration of type IIA string
theory.

In section 5, we comment on the future directions.

\section{$SU(N_c) \times SU(N_c')$ with $N_f$- and $N_f'$-fund. and bifund.}

\subsection{Electric theory}

The type IIA supersymmetric electric
brane configuration \cite{BH,BHKL,AT97,Ahn07-3} corresponding to 
${\cal N}=1$ $SU(N_c) \times SU(N_c')$ gauge theory  with  
$N_f$-fundamental flavors $Q, \widetilde{Q}$,
$N_f'$-fundamental flavors $Q', \widetilde{Q}'$
and bifundamentals $X, \widetilde{X}$
can be described as one middle NS5-brane, two
NS5'-branes, 
$N_c$- and $N_c'$-D4-branes, and $N_f$- and 
$N_f'$-D6-branes. The $X$ is in the representation $\bf{(\Box,\Box)}$ while 
the $\widetilde{X}$ is in the representation $\bf{(\overline{\Box},
\overline{\Box})}$
under the gauge group. 
The quarks $Q$ and $\widetilde{Q}$ are in the representation 
$\bf{(\Box, 1)}$ and $\bf{(\overline{\Box}, 1)}$ 
respectively under the gauge group.
Similarly, 
the quarks $Q'$ and $\widetilde{Q}'$ are in the representation 
$\bf{(1, \Box)}$ and $\bf{(1, \overline{\Box})}$ 
respectively under the gauge group.
The mass terms for each fundamental quarks 
can be added by displacing each D6-branes along 
\bea
v \equiv x^4 + i x^5
\nonu
\eea
direction leading to their coordinates 
$v = + 
v_{D6_{-\theta}}(+v_{D6_{-\theta'}})$ respectively  
while the quartic terms for each fundamental quarks 
can be added also by rotating each D6-branes
by an angle 
$-\theta(-\theta')$ in $(w,v)$-plane respectively.
Here we define the complex coordinate $w$ as 
\bea
w \equiv x^8 + i x^9.
\nonu 
\eea 
Then, in the electric gauge theory, the general 
superpotential is
given by
\bea
W_{elec} & = & \frac{\alpha}{2} \tr (Q \widetilde{Q})^2 - m \tr Q
\widetilde{Q} 
 + \frac{\alpha'}{2} \tr (Q' \widetilde{Q}')^2 - m' \tr Q'
\widetilde{Q}' \nonu \\
& + & \left[ -\frac{\beta}{2}   \tr (X \widetilde{X})^{2} + m_X
\tr X \widetilde{X}\right]
\label{elesuperpotential}
\eea 
where
$ \alpha \equiv \frac{\tan \theta}{\Lambda}$ and $  
m \equiv \frac{v_{D6_{-\theta}}}{2\pi \ell_s^2}$ for $N_f$ D6-branes and similarly 
$ \alpha' \equiv \frac{\tan \theta'}{\Lambda'}$ and $
m' \equiv \frac{v_{D6_{-\theta'}}}{2\pi \ell_s^2}$ for $N_f'$ D6-branes.
The last two terms of (\ref{elesuperpotential}) 
are due to the rotation angles $\omega_L$ and
$\omega_R$  
of two NS5'-branes in $(w,v)$-plane where
$\beta \equiv (\tan \omega_L + \tan \omega_R)$ 
and the relative displacement of two color D4-branes where the mass for the
bifundamentals $m_X
\equiv \frac{v_{NS5'}}{2\pi \ell_s^2}$
is the distance of D4-branes 
along the $v$-direction.
We focus on the particular limit 
$\beta, m_X \rightarrow 0$.

Then the ${\cal N}=1$ supersymmetric electric brane
configuration for the superpotential 
(\ref{elesuperpotential}) 
in type IIA string theory is given as follows and let us draw this
brane structure in
Figure 1 explicitly:

$\bullet$
One middle NS5-brane in $(012345)$ directions 

$\bullet$ 
Two NS5'-branes in  $(012389)$ directions 

$\bullet$
$N_f$ $D6_{-\theta}$-branes in (01237)
directions and
two other directions in $(v,w)$-plane

$\bullet$
$N_f'$ 
$D6_{-\theta'}$-branes in (01237)
directions and
two other directions in $(v,w)$-plane

$\bullet$
$N_c$- and $N_c'$-color D4-branes in $(01236)$ directions   

\begin{figure}[ht]
   \epsfxsize=4.0in 
\centerline{\epsffile{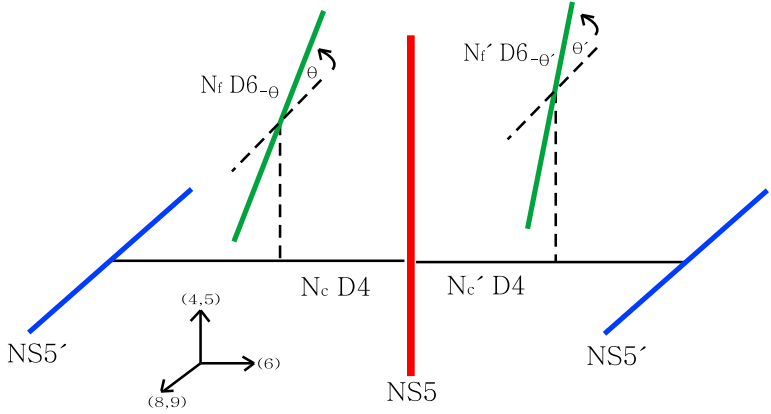}}
   \caption[FIG. \arabic{figure}.]{ 
The  ${\cal N}=1$ supersymmetric 
electric brane configuration for the gauge group $SU(N_c) \times SU(N_c')$ 
with bifundamentals $X, \widetilde{X}$ 
and fundamentals $Q, \widetilde{Q}, Q', \widetilde{Q}'$. 
A 
rotation of $N_f(N_f')$ D6-branes in $(w,v)$-plane
corresponds to 
a quartic term for the fundamentals $Q, \widetilde{Q}(Q', \widetilde{Q}')$ while 
a displacement of $N_f(N_f')$ D6-branes in $+v$ direction corresponds to a
mass term for the fundamentals $Q, \widetilde{Q}(Q', \widetilde{Q}')$.
}
\end{figure}

\subsection{Magnetic theory }

The left NS5'-brane 
starts out with linking number $l_e=-\frac{N_f'}{2} + N_c$
and after duality 
this left NS5'-brane ends up with linking number 
$l_m = \frac{N_f'}{2} -\widetilde{N}_c'+N_f$.
In general, when the $N_f$ $D6_{-\theta}$-branes meet 
the middle NS5-brane during the dual process,
the new D4-branes are created because they are not parallel.
However, we consider only the particular brane motion where
$N_f$ $D6_{-\theta}$-branes meet 
the middle NS5-brane with no angles. In other words, 
the  $D6_{-\theta}$-branes become $D6_{-\frac{\pi}{2}}$-branes 
when they meet with the middle NS5-brane instantaneously and then
after that
they come back to the original  $D6_{-\theta}$-branes.
Therefore, in this dual process, there is no creation of D4-branes.
That is the reason for the $N_f$ factor in the $l_m$, not $2N_f$.
Then it turns out that the dual color number $\widetilde{N}_c'$
is given by $\widetilde{N}_c' = N_f+N_f'-N_c$. 

What about the other dual color number?
Note that we take the Seiberg dual for both gauge group factors.
The right NS5'-brane 
starts out with linking number $l_e=\frac{N_f}{2} - N_c'$
and after duality 
this right NS5'-brane ends up with linking number 
$l_m = -\frac{N_f}{2} +\widetilde{N}_c-N_f'$.
In general, when the $N_f'$ $D6_{-\theta'}$-branes meet 
the middle NS5-brane during the dual process,
the new D4-branes are created because they are not parallel.
However, we consider only the particular brane motion where
$N_f'$ $D6_{-\theta'}$-branes meet 
the middle NS5-brane with no angles. In other words, 
the  $D6_{-\theta'}$-branes become $D6_{-\frac{\pi}{2}}$-branes 
when they meet with the middle NS5-brane instantaneously and after that 
they come back to the original  $D6_{-\theta'}$-branes.
Therefore, in this dual process, there is no creation of D4-branes.
That is the reason for the $N_f'$ factor in the $l_m$, not $2N_f'$.
Then it turns out that the dual color number $\widetilde{N}_c$
is given by $\widetilde{N}_c = N_f'+N_f-N_c'$. 
Finally, one has the following dual color numbers   
\bea
\widetilde{N}_c = N_f'+N_f-N_c', \qquad \widetilde{N}_c' = N_f+N_f'-N_c.
\nonu
\eea

The low energy theory on the color D4-branes 
has $SU(\widetilde{N}_c) \times SU(\widetilde{N}_c')$ gauge group and  
$N_f$-fundamental dual quarks $q', \widetilde{q}'$
coming from 4-4 strings connecting between the color 
$\widetilde{N}_c'$ D4-branes and
$N_f$ flavor D4-branes, 
$N_f'$-fundamental dual quarks $q, \widetilde{q}$
coming from 4-4 strings connecting between the color $\widetilde{N}_c$ D4-branes and
$N_f'$ flavor D4-branes
as well as $Y, \widetilde{Y}$ and various gauge singlets.
The $Y$ is in the representation $\bf{(\Box,\Box)}$ while 
the $\widetilde{Y}$ is in the representation $\bf{(\overline{\Box},
\overline{\Box})}$
under the dual gauge group. 
The $N_f'$ flavors $q$ and $\widetilde{q}$ are in the representation 
$\bf{(\Box, 1)}$ and $\bf{(\overline{\Box}, 1)}$ 
respectively under the gauge group and 
 in the representation 
$\bf{(\overline{\Box}, 1)}$ and $\bf{(1, \overline{\Box})}$ 
respectively under the flavor group $SU(N_f')_L \times SU(N_f')_R$.
Similarly, 
the $N_f$ flavors $q'$ and $\widetilde{q}'$ are in the representation 
$\bf{(1, \Box)}$ and $\bf{(1, \overline{\Box})}$ 
respectively under the gauge group
and 
 in the representation 
$\bf{(\overline{\Box}, 1)}$ and $\bf{(1, \overline{\Box})}$ 
respectively under the flavor group $SU(N_f)_L \times SU(N_f)_R$.
In particular, a magnetic meson field 
\bea
M_0 \equiv Q \widetilde{Q}
\nonu
\eea
is $N_f \times N_f$ matrix and comes from 
4-4 strings of $N_f$ flavor D4-branes while
a magnetic meson field 
\bea
M'_0 \equiv Q' \widetilde{Q}'
\nonu
\eea
is $N_f' \times N_f'$ matrix and comes from 
4-4 strings of $N_f'$ flavor D4-branes.
Then the most general magnetic superpotential 
is given by  
\bea
W_{dual} & = & \left[ (Y \widetilde{Y})^2+ Y \widetilde{Y}+  
M_0 q' \widetilde{Y} Y \widetilde{q}' + 
M_0' \widetilde{q} Y \widetilde{Y}  q
+ M_1 q' \widetilde{q}' + M_1' \widetilde{q} q + P q \widetilde{Y} q' + 
 \widetilde{P} \widetilde{q}' Y \widetilde{q} \right] 
\nonu \\
& + & \frac{\alpha}{2} \tr M_0^2 - m M_0 + \frac{\alpha'}{2} 
\tr {M_0'}^2 - m' M_0'
\label{poten}
\eea
where the mesons are
$
M_1 \equiv Q \widetilde{X} X \widetilde{Q},  
M_1'  \equiv Q'
\widetilde{X} X \widetilde{Q}', 
P  \equiv  Q \widetilde{X} Q'$ and  
$\widetilde{P} \equiv \widetilde{Q} X \widetilde{Q}'$.
The expression in the first line of 
(\ref{poten}) was found in \cite{ILS,BIWW} in the gauge theory side 
already.
Note that the mesons of the $SU(N_c)$ group  couple to the dual quarks
of $SU(\widetilde{N}_c')$ and 
the mesons of the $SU(N_c')$ group  couple to the dual quarks
of $SU(\widetilde{N}_c)$ \footnote{As suggested in \cite{BH}, one can
dualize each gauge group independently of the other. For example, one
dualizes the second gauge group factor by moving the middle NS5-brane
to the right of the right NS5'-brane, like as in \cite{Ahn07-3}. Then
the dual gauge group is given by $SU(N_c) \times SU(\widetilde{n}_c' =
N_f'+N_c-N_c')$ with corresponding superpotential for the 
dual matters. Now we interchange the NS5'-branes and 
$D6_{-\theta,-\theta'}$-branes each other and obtain the next dual gauge
group
 $SU(\widetilde{n}_c=2N_f'+N_f-N_c') \times SU(\widetilde{n}_c')$ with
dual matters. Finally, we move the left NS5'-brane and
$D6_{-\theta}$-branes(in the electric
theory) to the right of the middle NS5-brane and obtain the final dual
gauge group $SU(\widetilde{N}_c=\widetilde{n}_c) \times
SU(\widetilde{N}_c'=2N_f+N_f'-N_c)$
with the superpotential (\ref{poten}). We also obtain the same dual
gauge theory
if we start with $SU(N_c)$ dualization first and then $SU(N_c')$
dualization and finally end up with $SU(n_c)$ dualization. }.

As we explained before, 
our particular brane motion during the dual process does not produce
any D4-branes 
when the $N_f$ $D6_{-\theta}$-branes meet the middle NS5-brane. This  
implies that there is no $M_1$ term in the above superpotential
(\ref{poten}).
The meson 
$M_1$ originates from  $SU(N_c)$ chiral mesons
$Q\widetilde{Q}$  when one
dualizes the first gauge group factor first by moving the middle NS5-brane
to the left of the left NS5'-brane. That is, the fluctuations of
strings stretching between the $N_f$ ``flavor'' D4-branes correspond
to this meson field. 
The superpotential contains the cubic term between this
meson field and dual quarks. After two additional dual procedures, 
$SU(N_c')$ and $SU(\widetilde{n}_c)$, this cubic term arises as 
$M_1$-term in (\ref{poten}) where $M_1$ has an
extra $\widetilde{X} X$ fields besides $Q \widetilde{Q}$, 
due to the further $SU(N_c')$-dualization.

Similarly, 
because there is no creation of D4-branes
when 
the $N_f'$ $D6_{-\theta'}$-branes 
meet the middle NS5-brane, 
there is no $M_1'$ term in the above superpotential
(\ref{poten}) also.
The meson 
$M_1'$ originates from  $SU(N_c')$ chiral mesons
$Q'\widetilde{Q}'$  when one
dualizes the second gauge group factor first by moving the middle NS5-brane
to the right of the right NS5'-brane. That is, the
strings stretching between the $N_f'$ ``flavor'' D4-branes provides
this meson. 
The superpotential in
\cite{Ahn07-3} contains the cubic term between this
meson field and dual quarks. After two additional dual procedures, 
$SU(N_c)$ and $SU(\widetilde{n}_c')$ observed in the footnote 1, 
this cubic term arises as 
$M_1'$-term in (\ref{poten}) where $M_1'$ has 
extra $\widetilde{X} X$ fields besides $Q' \widetilde{Q}'$, 
due to the further $SU(N_c)$-dualization.

Furthermore, 
we do not see any $P$- or $\widetilde{P}$-dependent terms in the
superpotential (\ref{poten}) when the $N_f$ $D6_{-\theta}$-branes, the 
$N_f$ $D6_{-\theta'}$-branes and the middle NS5-brane during the dual
process
meet each other with no angles.
These mesons 
$P$ and $\widetilde{P}$ originate from  $SU(N_c')$ chiral mesons
$\widetilde{X}Q'$ and $X\widetilde{Q}'$ \cite{Ahn07-3} when one
dualizes the second gauge group factor first by moving the middle NS5-brane
to the right of the right NS5'-brane, as in footnote 1. That is, the
strings stretching between the $N_f'$ flavor D4-branes and $N_c$
color D4-branes give rise to these $N_f'$ $SU(N_c)$ fundamentals and 
$N_f'$ $SU(N_c)$ antifundamentals. The superpotential in
\cite{Ahn07-3} contains the cubic term between dual bifundamental, these
meson fields and dual quarks. After two additional dual procedures, 
$SU(N_c)$ and $SU(\widetilde{n}_c')$, these cubic terms arise as $P$
and $\widetilde{P}$-term in (\ref{poten}) where there exist  extra 
quarks $q$ and $\widetilde{q}$ while $P$ and $\widetilde{P}$ have 
extra $Q$ and $\widetilde{Q}$ fields, 
due to the further $SU(N_c)$-dualization. 

All these features for selecting the wanted gauge singlet during the
dual process has occurred also in the different gauge theories 
\cite{Ahn07,Ahn07-1} where the gauge group is a single gauge group
with the presence of O6-plane and in these cases the brane 
does not move independently due to the O6-plane.
Then the reduced magnetic superpotential in our case 
with the limit $\beta, m_X
\rightarrow 0$ 
is given by 
\bea
W_{dual}  = \left[
M_0 q' \widetilde{Y} Y \widetilde{q}' 
 +  \frac{\alpha}{2} \tr M_0^2 - m M_0  \right] 
+  \left[  M_0' \widetilde{q} Y \widetilde{Y}  q + \frac{\alpha'}{2} \tr
 {M_0'}^2 - 
m' M_0'\right].
\label{dualW1}
\eea

For the supersymmetric vacua, one can compute the F-term equations for
this superpotential (\ref{dualW1}) 
and the expectation values for $M_0, M'_0, q' \widetilde{Y} Y \widetilde{q}'$ 
and $\widetilde{q} Y \widetilde{Y}  q$ are obtained.
The F-term equations for $M_0, q', \widetilde{q}', M_0',
q, \widetilde{q}, Y$ and $\widetilde{Y}$ are 
\bea
q' \widetilde{Y} Y \widetilde{q}' -m + \alpha M_0  & = & 0, \qquad
\widetilde{Y} (Y \widetilde{q}' M_0) =0, \qquad 
(M_0 q' \widetilde{Y}) Y =0, \nonu \\
 \widetilde{q} Y \widetilde{Y} q -m' + \alpha' M_0' & = & 0, \qquad
(M_0' \widetilde{q} Y) \widetilde{Y} =0, \qquad 
Y (\widetilde{Y} q M_0') =0, \nonu \\
\widetilde{q}' (M_0 q' \widetilde{Y}) + (\widetilde{Y} q M_0')
\widetilde{q} & = & 0, \qquad 
(Y \widetilde{q}' M_0) q' + q (M_0' \widetilde{q} Y) =0.
\label{Fterm}
\eea
The seventh and eighth equations of (\ref{Fterm}) are satisfied if the  
the second, third, fifth and sixth equations of (\ref{Fterm}) hold:
$
Y \widetilde{q}' M_0 = M_0 q' \widetilde{Y} =
M_0' \widetilde{q} Y =\widetilde{Y} q M_0' =0$. 
We present the magnetic brane configuration in Figure 2. 
Depending on the values of the masses $m$ and $m'$, there exist other
three possibilities: the $v$ coordinates of $D6_{-\theta}$-branes and 
$D6_{-\theta'}$-branes  are classified as
$v=(+,-), (-,+)$ and $(-,-)$. 
Of course, if we put an orientifold 6-plane at the origin, then this
figure reduces to the one in \cite{Ahn07} where the matter contents
and superpotential should be preserved under the O6-plane.  

\begin{figure}[ht]
   \epsfxsize=3.5in 
\centerline{\epsffile{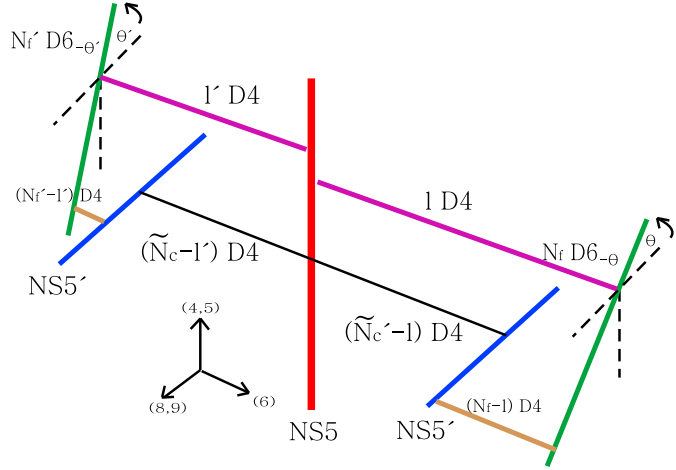}}
   \caption[FIG. \arabic{figure}.]{ 
The  ${\cal N}=1$ supersymmetric
magnetic brane configuration corresponding to Figure 1 with 
a splitting and a reconnection 
between D4-branes when the gravitational potential of
the NS5-brane is ignored. 
The $N_f$ flavor D4-branes connecting between
$D6_{-\theta}$-branes and NS5'-brane are splitting into $(N_f-l)$- and
$l$- D4-branes while
the $N_f'$ flavor D4-branes connecting between
$D6_{-\theta'}$-branes and NS5'-brane are splitting into $(N_f'-l')$- and
$l'$- D4-branes. 
The $v$ and $w$ coordinates of each these D4-branes are related to each
other through the deformation parameters $\theta$ and $\theta'$ 
respectively.  }
\end{figure}

The theory has many nonsupersymmetric meta-stable ground states and 
when we rescale the meson fields as
\bea
M_0 = h \Lambda \Phi_0 \qquad 
\mbox{and} \qquad
M'_0 = h' \Lambda' \Phi'_0,
\nonu
\eea
then the Kahler potential for $\Phi_0$ and $\Phi'_0$ 
is canonical and the magnetic
quarks are canonical near the origin of field space \cite{ISS}.
Then the magnetic superpotential (\ref{dualW1}) can be rewritten as
\bea
W_{mag} = \left[ h \Phi_0  q'   \widetilde{Y} Y \widetilde{q}' 
 +  
\frac{h^2 \mu_{\phi}}{2} \tr \Phi_0^2- h \mu^2 \tr \Phi_0 \right] + 
\left[ h' \Phi_0'     \widetilde{q} Y \widetilde{Y} q 
 +  
\frac{{h'}^2 \mu_{\phi}'}{2} \tr {\Phi_0'}^2- 
h' {\mu'}^2 \tr \Phi_0'\right]
\nonu
\eea
where
$
\mu^2 = m \Lambda, {\mu'}^2 =m' \Lambda'$ and  
$\mu_{\phi} = \alpha \Lambda^2, \mu_{\phi}' = \alpha' {\Lambda'}^2$.

Now one splits 
the $(N_f-l) \times (N_f-l)$
block  at the lower right corner of $h\Phi_0$ and $q' \widetilde{Y} Y 
\widetilde{q}'$ 
into blocks of 
size $n$ and $(N_f-l-n)$ and 
 one decomposes 
the $(N_f'-l') \times (N_f'-l')$
block  at the lower right corner of $h' \Phi'_0$ and $\widetilde{q} Y 
\widetilde{Y}  q$ 
into blocks of 
size $n'$ and $(N_f'-l'-n')$
as follows \cite{GK0710}:
\bea
h\Phi_0 & = &  \left(
\begin{array}{ccc}
0_l & 0 & 0  \\
0 & h \Phi_n & 0 \\
0 & 0 & \frac{\mu^2}{\mu_{\phi}} {\bf 1}_{N_f-l-n}
\end{array}
\right), \qquad
h' \Phi_0'  =   \left(
\begin{array}{ccc}
0_{l'} & 0 & 0  \\
0 & h' \Phi_{n'}' & 0 \\
0 & 0 & \frac{{\mu'}^2}{\mu_{\phi}'} {\bf 1}_{N_f'-l'-n'}
\end{array}
\right), \nonu \\
q'  \widetilde{Y} Y \widetilde{q}'  & = & \left(
\begin{array}{ccc}
\mu^2 {\bf 1}_l & 0 & 0  \\
0 & { \varphi}' \widetilde{y} y  \widetilde{\varphi}'  &  0 \\
0 & 0 & 0_{N_f-l-n}
\end{array}
\right), \qquad
\widetilde{q} Y \widetilde{Y} q = \left(
\begin{array}{ccc}
{\mu'}^2 {\bf 1}_{l'} & 0 & 0  \\
0 & \widetilde{\varphi} y \widetilde{y}  \varphi  &  0 \\
0 & 0 & 0_{N_f'-l'-n'}
\end{array}
\right).
\nonu
\eea
Here $\varphi'$ and $\widetilde{\varphi}'$ are $n \times (\widetilde{N}_c'-l)$
dimensional matrices and correspond to $n$ flavors of fundamentals of
the gauge group $SU(\widetilde{N}_c'-l)$ which is unbroken and 
$\varphi$ and $\widetilde{\varphi}$ are $n' \times (\widetilde{N}_c-l')$
dimensional matrices and correspond to $n'$ flavors of fundamentals of
the gauge group $SU(\widetilde{N}_c-l')$ which is unbroken.
In the brane configuration shown in Figure 3, 
$\varphi'$ and $\widetilde{\varphi}'$ correspond to 
fundamental strings connecting the $n$ flavor D4-branes and
$(\widetilde{N}_c'-l)$
color D4-branes and 
$\varphi$ and $\widetilde{\varphi}$ correspond to 
fundamental strings connecting the $n'$ flavor D4-branes and
$(\widetilde{N}_c-l')$
color D4-branes.
The $\Phi_n$ and ${ \varphi}' \widetilde{y} y 
\widetilde{\varphi}'$
are $n \times n$ matrices while 
$\Phi'_{n'}$ and $\widetilde{ \varphi} 
y \widetilde{y} \varphi$
are $n' \times n'$ matrices.
The supersymmetric ground state corresponds to
\bea
h\Phi_n= \frac{\mu^2}{\mu_{\phi}} {\bf 1}_{n}, 
\qquad \varphi' \widetilde{y} =0=y \widetilde{\varphi}' \qquad 
\mbox{and} \qquad 
h'\Phi'_n= \frac{{\mu'}^2}{\mu_{\phi}'} {\bf 1}_{n'}, 
\qquad
\widetilde{y} \varphi =0=\widetilde{\varphi} y.
\nonu 
\eea 

Now the full one loop potential, by combining the superpotential 
and the vacuum expectation values for the fields, 
takes the form
\bea
V  & = &  
|h \Phi_n \varphi' \widetilde{y}|^2   
+  |h  y \widetilde{\varphi}' \Phi_n|^2
  +  
| h \varphi'  \widetilde{y} y \widetilde{\varphi}'-h \mu^2 {\bf 1}_{n} + 
h^2 \mu_{\phi} \Phi_n|^2 + b |h^2 \mu|^2 \tr \Phi_n^{\dagger} \Phi_n
\nonu \\ 
& + & |h' \Phi_{n'}'  \widetilde{\varphi} y |^2   
+  |h'  \widetilde{y} \varphi \Phi_{n'}' |^2
  +  
|   h' \widetilde{\varphi} y \widetilde{y} \varphi -h' {\mu'}^2 {\bf 1}_{n'} + 
{h'}^2 \mu_{\phi}' \Phi_{n'}'|^2 + b' |{h'}^2 \mu'|^2 
\tr {\Phi'}^{\dagger}_{n'} 
\Phi_{n'}'
\nonu
\eea
where $b = \frac{(\ln 4-1)}{8\pi^2} \widetilde{N}_c$ and 
$b' = \frac{(\ln 4-1)}{8\pi^2} \widetilde{N}_c'$ \cite{ISS}.
Differentiating this potential with respect to 
$\Phi_n^{\dagger}$ and ${\Phi'}^{\dagger}_{n'}$ and putting 
$\varphi' \widetilde{y} =0 =  y \widetilde{\varphi}'$ and 
$ \widetilde{\varphi} y=0 =  \widetilde{y} \varphi$ \cite{GK0710}, 
one obtains, using the methods given in 
\cite{Ahn07-11,Ahn08-1,Ahn08-1two,Ahn08-2,Ahn08-3},
\bea
h \Phi_n 
& \simeq &  \frac{ \mu_{\phi}}{b }
{\bf 1}_n   \qquad  \mbox{or} \qquad
M_n \simeq \frac{\alpha \Lambda^3}{\widetilde{N}_c} {\bf 1}_{n}, 
\nonu \\
h' \Phi_{n'}' 
& \simeq &  \frac{ \mu_{\phi}'}{b' }
{\bf 1}_{n'}  \qquad  \mbox{or} \qquad
M'_{n'} \simeq \frac{\alpha' {\Lambda'}^3}{\widetilde{N}_c'} {\bf
  1}_{n'} 
\nonu
\eea
corresponding to the $w$ coordinates of $n$ curved flavor D4-branes between 
the $D6_{-\theta}$-branes and the NS5'-brane and  
the $w$ coordinates of $n'$ curved flavor D4-branes between 
the $D6_{-\theta'}$-branes and the NS5'-brane respectively.

\begin{figure}[ht]
   \epsfxsize=3.5in 
\centerline{\epsffile{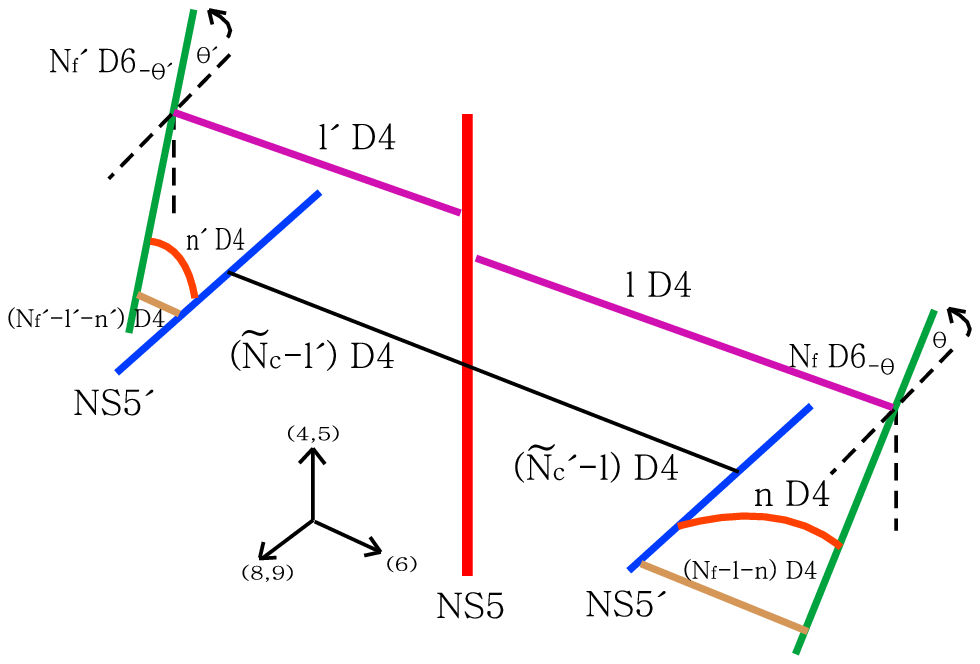}}
   \caption[FIG. \arabic{figure}.]{ 
The  nonsupersymmetric meta-stable
magnetic brane configuration corresponding to Figure 1 with 
a misalignment between D4-branes when the gravitational potential of
the NS5-brane is considered. 
The $(N_f-l)$ flavor D4-branes in Figure 2 connecting between
$D6_{-\theta}$-branes and NS5'-brane are further splitting into $(N_f-l-n)$- and
$n$-curved D4-branes while
the $(N_f'-l')$ flavor D4-branes in Figure 2 connecting between
$D6_{-\theta'}$-branes and NS5'-brane are further 
splitting into $(N_f'-l'-n')$- and
$n'$-curved D4-branes. 
When there are two multiple NS5'-branes, 
further distributions of D4-branes arise along the $v$ direction.  }
\end{figure}

\subsection{Higher order superpotential for bifundamentals}

In general, there are also different meson fields 
$M_j = Q (\widetilde{X} X)^j \widetilde{Q}, M'_j = Q' (\widetilde{X}
X)^j \widetilde{Q}',
P_r = Q (\widetilde{X}
X)^{r-1} \widetilde{X} Q' $ and $ 
\widetilde{P}_r = \widetilde{Q} X (\widetilde{X}
X)^{r-1}  \widetilde{Q}'$ where $j=0, 1, \cdots,
k'-1$ and $r=1, \cdots, k'$ for higher order superpotential for
bifundamental with general rotation angles of
two $k'$ NS5'-branes \cite{ILS,BIWW,BH}.
The magnetic superpotential contains the interaction between these
meson fields with $q, \widetilde{q}, q', \widetilde{q}', Y$ and
$\widetilde{Y}$ 
\cite{ILS}
as well as the higher order term for dual bifundamental 
$(Y \widetilde{Y})^{k'+1}$ and mass term for dual 
bifundamental $Y \widetilde{Y}$ 
which will vanish for small
$\beta, m_X$ limit. 
When the
$N_f$ $D6_{-\theta}$-branes and the extra right $(k'-1)$ NS5'-branes 
do not create the D4-branes
and $N_f'$ $D6_{-\theta'}$-branes and the extra left $(k'-1)$ NS5'-branes
also do not
create the D4-branes,  the extra meson fields $M_j, M_j', P_r$ and $
\widetilde{P}_r$ where $j \neq 0$ 
do not appear in the
magnetic superpotential. 

Then 
the $k'$-dependent magnetic superpotential 
with the limit $\beta, m_X
\rightarrow 0$ 
is given by 
\bea
W_{dual}  = \left[
M_0 q' (\widetilde{Y} Y)^{k'} \widetilde{q}' 
 +  \frac{\alpha}{2} \tr M_0^2 - m M_0  \right] 
+  \left[  M_0' \widetilde{q} (Y \widetilde{Y})^{k'}  q + \frac{\alpha'}{2} \tr
 {M_0'}^2 - 
m' M_0'\right].
\nonu
\eea  
Then the analysis for the previous single NS5'-brane case can be
performed in this case also. 
One deforms the generalized Figure 3, where there are multiple NS5'-branes, 
by displacing the multiple
$D6_{-\theta,-\theta'}$-branes and $NS5'$-branes 
along $v$ direction, as in \cite{Ahn08-2,Ahn08-3}.
Then the $n$ curved flavor D4-branes attached to them(as well as other
D4-branes) are displaced
also as $k'$ different $n_j$'s connecting between 
$D6_{-\theta,j}$-brane and
the right $NS5_{j}'$-brane$(j=1, 2, \cdots, k')$.
Similarly,
the $n'$ curved flavor D4-branes attached to them(as well as other
D4-branes) are displaced
also as $k'$ different $n'_{j}$'s connecting between 
$D6_{-\theta',j}$-brane and
the left $NS5_{j}'$-brane.
One can consider the particular case $N_f=k'=N_f'$.
When we rescale the submeson fields as
$M_{j} = h \Lambda \Phi_{j} $ and $M'_{j} = h' \Lambda' \Phi'_{j}$ \cite{Ahn08-2},
then the Kahler potential for $\Phi_{j}$ and $\Phi'_{j}$ 
is canonical and the magnetic
quarks $q_j, \widetilde{q}_j$ and $q'_{j}, \widetilde{q}'_{j}$ 
are canonical near the origin of field space \cite{ISS}.
Then the magnetic superpotential  
can be rewritten in terms of $\Phi_j, q_j, \widetilde{q}_j, 
\Phi'_{j}, q'_{j}, \widetilde{q}'_{j}, 
Y_j, Y_{j'}, \widetilde{Y}_{j}$ and $\widetilde{Y}_{j'}$.
In order to see the nonsupersymmetric
meta-stable ground states, one can   
split  some of the components of $h\Phi_0, q' (\widetilde{Y} Y)^{k'} 
\widetilde{q}', h' \Phi'_0$ and $\widetilde{q} (Y \widetilde{Y})^{k'}
q$ as usual.
The supersymmetric ground state corresponds to
\bea
h\Phi_{n_j}  & = & \frac{\mu_j^2}{\mu_{\phi}} {\bf 1}_{n_j}, 
\qquad 
\varphi_{n_j}' (\widetilde{y}_{n_j} y_{n_j} )^{\frac{k'}{2}}  =0=   
(\widetilde{y}_{n_j} y_{n_j} )^{\frac{k'}{2}}
\widetilde{\varphi}_{n_j}' \qquad \mbox{and} \nonu \\ 
h'\Phi'_{n_{j}'} & = & \frac{{\mu_{j}'}^2}{\mu_{\phi}'} {\bf 1}_{n_{j}'}, 
\qquad (y_{n_j'} \widetilde{y}_{n_j'})^{\frac{k'}{2}} \varphi_{n'_{j}} =
0=\widetilde{\varphi}_{n'_{j}} 
(y_{n_j'} \widetilde{y}_{n_j'})^{\frac{k'}{2}}
\nonu
\eea
when $k'$ is even.
For $k'$ odd, one gets similar supersymmetric ground states. 
The full one loop potential can be written similarly
and the local nonzero stable point arises as
\bea
h \Phi_{n_j} 
& \simeq &  \frac{ \mu_{\phi}}{b_j }
{\bf 1}_{n_j}   \qquad  \mbox{and} \qquad
h' \Phi_{n_{j}'}' 
 \simeq   \frac{ \mu_{\phi}'}{b_{j}' }
{\bf 1}_{n_{j}'}  
\nonu
\eea  
corresponding to the $w$ coordinates of $n_j$ curved flavor D4-branes between 
the $D6_{-\theta,j}$-branes and the $NS5_{R,j}'$-brane and  
the $w$ coordinates of $n_{j}'$ curved flavor D4-branes between 
the $D6_{-\theta',j}$-branes and the $NS5_{L,j}'$-brane respectively.

Therefore, the meta-stable states, for fixed $k'$ which is related to 
the order of the bifundamental field in the superpotential  and $\theta$ and
$\theta'$ which are 
deformation parameters by rotation angles of $D6_{-\theta, -\theta'}$-branes, 
are classified by the number of various D4-branes and the positions of
multiple $D6_{-\theta, -\theta'}$-branes and 
NS5'-branes \cite{Ahn08-2,Ahn08-3}: 
\bea
(N_{c,j}, N_{f,j}, l_j, n_j, N'_{c,j}, N_{f,j}', l_j', n_j') \qquad 
\mbox{and} \qquad
(v_{D6_{-\theta,j}}, v_{D6_{-\theta',j}}, v_{NS5_{L,j}'},
v_{NS5_{R,j}'})
\nonu
\eea
where $j=1, 2, \cdots, k'$.
The description of 
other range for the $N_f, N_f'$ and $k'$ can be analyzed similarly.

\section{
$SO(2N_c) \times Sp(N_c')$ with $2N_f$-vectors,  $2N_f'$-fund.  
and bifund.}

Let us add orientifold 4-plane to the previous brane configuration.

\subsection{Electric theory}

The type IIA supersymmetric electric
brane configuration \cite{Ahn07-2,Tatar,Ahn97} corresponding to 
${\cal N}=1$ $SO(2N_c) \times Sp(N_c')$ gauge theory  with  
$2N_f$-vectors $Q$,
$N_f'$-fundamental flavors $Q'$
and bifundamental $X$
can be described as one middle NS5-brane, two
NS5'-branes, 
$2N_c$- and $2N_c'$-D4-branes, and $2N_f$- and   
$2N_f'$-D6-branes and  $O4^{\pm}$-planes. 
The mass terms can be achieved by displacing each D6-branes along $ \pm v$
direction leading to their coordinates $v = 
\pm v_{D6_{-\theta}}(\pm v_{D6_{-\theta'}})$  respectively
while the quartic terms for the quarks 
can be obtained by rotating the D6-branes
by an angle 
$-\theta(-\theta')$ in $(w,v)$-plane respectively. 
Then, in the electric gauge theory, the general superpotential is
given by
\bea
W_{elec} & = & \frac{\alpha}{2} \tr (Q Q)^2 - m \tr Q Q +
 \frac{\alpha'}{2} \tr (Q' Q')^2 - m' \tr Q'
Q' \nonu \\
  &+& \left[  -
\frac{\beta}{2} \tr (X X)^{2} + m_X \tr X X\right].
\label{esuperpotent}
\eea 
The last two terms of (\ref{esuperpotent}) 
are due to the rotation angle $\omega$ of NS5'-branes where
$\beta=\tan \omega $ 
and the relative displacement of D4-branes where the mass for the
bifundamental $m_X =\frac{v_{NS5'}}{2\pi \ell_s^2}$ 
is the distance of D4-branes in $v$ direction.
We focus on the limit 
$\beta, m_X \rightarrow 0$.

Let us summarize the ${\cal N}=1$ supersymmetric electric brane
configuration for the superpotential 
(\ref{esuperpotent}) 
in type IIA string theory 
and we do not draw this here but it is given by 
Figure 1 with the appearance of mirrors for an orientifold 4-plane.

$\bullet$
One middle NS5-brane in $(012345)$ directions  

$\bullet$ 
Two NS5'-branes in  $(012389)$ directions 

$\bullet$
$2N_f$ $D6_{-\theta}$-branes in (01237)
directions and
two other directions in $(v,w)$-plane

$\bullet$
$2N_f'$ 
$D6_{-\theta'}$-branes in (01237)
directions and
two other directions in $(v,w)$-plane

$\bullet$
$2N_c$- and $2N_c'$-color D4-branes in $(01236)$ directions 

$\bullet$ $O4^{\pm}$-planes in (01236) directions 

\subsection{Magnetic theory }

The left NS5'-brane 
starts out with linking number $l_e=-\frac{(2N_f')}{2}-2 + 2N_c$
and after duality 
this left NS5'-brane ends up with linking number 
$l_m = \frac{(2N_f')}{2}-2 -2\widetilde{N}_c'+2N_f$.
We consider only the particular brane motion where
$2N_f$ $D6_{-\theta}$-branes meet 
the middle NS5-brane with no angles. In other words, 
the  $D6_{-\theta}$-branes become $D6_{-\frac{\pi}{2}}$-branes 
when they meet with NS5-brane instantaneously and then 
after that they come back to the original  $D6_{-\theta}$-branes.
Then it turns out that the dual color number $2\widetilde{N}_c'$
is given by $2\widetilde{N}_c' = 2N_f+2N_f'-2N_c$. 

The right NS5'-brane 
starts out with linking number $l_e=\frac{(2N_f)}{2}-2 - 2N_c'$
and after duality 
this right NS5'-brane ends up with linking number 
$l_m = -\frac{(2N_f)}{2}-2 +2\widetilde{N}_c-2N_f'$.
We consider only the particular brane motion where
$2N_f'$ $D6_{-\theta'}$-branes meet 
the middle NS5-brane with no angles. 
The  $D6_{-\theta'}$-branes become $D6_{-\frac{\pi}{2}}$-branes 
when they meet with NS5-brane instantaneously and then 
after that they come back to the original  $D6_{-\theta'}$-branes.
Then it turns out that the dual color number $2\widetilde{N}_c$
is given by $2\widetilde{N}_c = 2N_f'+2N_f-2N_c'$. 
Finally, one has the following dual color numbers   
\bea
2\widetilde{N}_c = 2N_f'+2N_f-2N_c', \qquad 
2\widetilde{N}_c'= 2N_f+ 2N_f'-2N_c.
\nonu
\eea

The low energy theory on the color D4-branes 
has $SO(2\widetilde{N}_c) \times Sp(\widetilde{N}_c')$ gauge group and  
$2N_f$-fundamental dual quarks $q'$
coming from 4-4 strings connecting between 
the color $2\widetilde{N}_c'$ D4-branes and
$2N_f$ flavor D4-branes, 
$2N_f'$-fundamental dual quarks $q$
coming from 4-4 strings connecting between the color 
$2\widetilde{N}_c$ D4-branes and
$2N_f'$ flavor D4-branes
as well as $Y$ and various gauge singlets.
The $2N_f'$ flavors $q$ and $2N_f$ flavors $q'$ are in the representation 
$\bf{(\Box, 1)}$ and $\bf{(1, \Box)}$ 
respectively under the gauge group and 
 in the representation 
$\bf{(1, \overline{\Box})}$ and $\bf{(\overline{\Box}, 1)}$ 
respectively under the flavor group $SU(2N_f) \times SU(2N_f')$.
In particular, a magnetic meson field 
\bea
M_0 \equiv Q Q
\nonu
\eea
is $2N_f \times 2N_f$ symmetric matrix and comes from 
4-4 strings of $2N_f$ flavor D4-branes while
a magnetic meson field 
\bea
M'_0 \equiv Q' Q'
\nonu
\eea
is $2N_f' \times 2N_f'$ antisymmetric matrix and comes from 
4-4 strings of $2N_f'$ flavor D4-branes.
Then the most general magnetic superpotential 
is given by  
\bea
W_{dual} & = & \left[  
(Y Y)^2 + Y Y + M_0 q' Y Y q' + 
M_0' q Y Y  q
+ M_1 q' q' + M_1' q q + P q Y q'  \right] 
\nonu \\
& + & \frac{\alpha}{2} \tr M_0^2 - m M_0 + 
\frac{\alpha'}{2} \tr {M_0'}^2 - m' M_0'
\label{dualW2}
\eea
where the mesons are
$
M_1 \equiv Q X X Q, 
M_1'  \equiv Q' X X Q'$ and $ 
P  \equiv  Q X Q'$.
The first line of (\ref{dualW2}) was appeared in \cite{ILS}.
Note that the mesons of the $SO(2N_c)$ group  couple to the dual quarks
of $Sp(\widetilde{N}_c')$ and 
the mesons of the $Sp(N_c')$ group  couple to the dual quarks
of $SO(2\widetilde{N}_c)$. 

One can
dualize each gauge group independently of the other. For example, one
dualizes the second gauge group factor by moving the middle NS5-brane
to the right of the right NS5'-brane, like as in \cite{Ahn07-2}. Then
the dual gauge group is given by $SO(2N_c) \times Sp(\widetilde{n}_c' =
N_f'+N_c-N_c'-2)$ with corresponding superpotential for the 
dual matters. Now we interchange the NS5'-branes and 
$D6_{-\theta,-\theta'}$-branes each other and obtain the dual gauge
group
 $SO(2\widetilde{n}_c=4N_f'+2N_f-2N_c') \times Sp(\widetilde{n}_c')$ with
dual matters. Finally, we move the left NS5'-brane and
$D6_{-\theta}$-branes 
in the electric
theory to the right of the middle NS5-brane and obtain the final dual
gauge group $SO(2\widetilde{N}_c=2\widetilde{n}_c) \times
Sp(\widetilde{N}_c'=2N_f+N_f'-N_c)$
with the superpotential (\ref{dualW2}) \footnote{As in unitary case of
previous section,
it does not matter the order of dualization. We obtain the same
dual gauge theory if we start with $SO(2N_c)$ dualization, then $Sp(N_c')$
dualization and end up with $SO(2\widetilde{n}_c)$ dualization.}.

As before, 
our particular brane motion during the dual process does not produce
any D4-branes 
when the $2N_f$ $D6_{-\theta}$-branes meet the middle NS5-brane. This  
implies that there is no $M_1$ term in the above superpotential
(\ref{dualW2})
 \footnote{The meson 
$M_1$ originates from  $SO(2N_c)$ chiral mesons
$Q Q$  when one
dualizes the first gauge group factor first. That is, the fluctuation
of the
strings stretching between the $2N_f$ ``flavor'' D4-branes provides this
meson field. 
The superpotential at this stage contains the cubic term between this
meson field and dual quarks. After two additional dual procedures, 
$Sp(N_c')$ and $SO(2\widetilde{n}_c)$, this cubic term arises as 
$M_1$-term in (\ref{dualW2}) where $M_1$ has an
extra $X X$ field-dependent factor besides $Q Q$, 
due to the further $Sp(N_c')$-dualization. }.
Similarly, 
when 
the $2N_f'$ $D6_{-\theta'}$-branes 
meet the middle NS5-brane, the fact that there is no creation of
D4-branes leads to the fact that
there is no $M_1'$ term in the above superpotential
(\ref{dualW2}) also
 \footnote{The meson 
$M_1'$ originates from  $Sp(N_c')$ chiral mesons
$Q' Q'$  when one
dualizes the second gauge group factor first. Then the fluctuations of
the
strings stretching between the $2N_f'$ ``flavor'' D4-branes gives this
meson field. 
The superpotential in
\cite{Ahn07-2} contains the cubic term between this
meson field and dual quarks. After two additional dual procedures, 
$SO(2N_c)$ and $Sp(\widetilde{n}_c')$, this cubic term arises as 
$M_1'$-term in (\ref{dualW2}) where $M_1'$ has 
extra $X X$ fields besides $Q' Q'$, 
due to the further $SO(2N_c)$-dualization. }.
Furthermore, 
we do not see any $P$-dependent term in the
superpotential (\ref{dualW2}) when the $2N_f$ $D6_{-\theta}$-branes, the 
$2N_f'$ $D6_{-\theta'}$-branes and the middle NS5-brane, during the dual
process,
meet each other with no angles \footnote{The meson 
$P$ originates from  $Sp(N_c')$ chiral meson
$X Q'$  \cite{Ahn07-2} when one
dualizes the second gauge group factor first. That is, the
strings stretching between the $2N_f'$ ``flavor'' D4-branes and $2N_c$
color D4-branes give rise to these $2N_f'$ $SO(2N_c)$ flavors. 
The superpotential contains the cubic term between dual bifundamental, this
meson field and dual quarks. After two additional dual procedures, 
$SO(2N_c)$ and $Sp(\widetilde{n}_c')$, this cubic term arises as
$P$-term 
in (\ref{dualW2}) where there exist  extra 
quarks $q$ while $P$ has an
extra $Q$ field, 
due to the further $SO(2N_c)$-dualization. }. 
Then the reduced magnetic superpotential in our case 
with the limit $\beta, m_X
\rightarrow 0$ 
is given by 
\bea
W_{dual}  = \left[
M_0 q' Y Y q' 
 +  \frac{\alpha}{2} \tr M_0^2 - m M_0  \right] 
+  \left[  M_0' q Y Y  q + \frac{\alpha'}{2} \tr
 {M_0'}^2 - 
m' M_0'\right].
\label{Dual}
\eea

For the supersymmetric vacua, one can compute the F-term equations for
this superpotential (\ref{Dual}) 
and the expectation values for $M_0, M'_0, q' Y Y q'$ 
and $q Y Y  q$ are obtained.
The F-term equations for $M_0, q', M_0',
q$ and $Y$ are 
\bea
q' Y Y q' -m + \alpha M_0  & = & 0, \qquad
Y (Y q' M_0) =0, \nonu \\
 q Y Y q -m' + \alpha' M_0' & = & 0, \qquad
(M_0' q Y) Y =0, \nonu \\
(Y q' M_0) q'  +  q (M_0' q Y) & = & 0.
\label{Fterm1}
\eea
The fifth equation of (\ref{Fterm1}) is satisfied if the  
second and fourth equations of (\ref{Fterm1}) hold:
$
Y q' M_0 = M_0' q Y =0$. 
One can read off the corresponding magnetic brane configuration by
adding O4-plane and mirrors from
the Figure 2.

The theory has many nonsupersymmetric meta-stable ground states by
requiring the IR free region for each gauge group factors \cite{Ahn06-1} and 
when we rescale the meson fields as
$
M_0 = h \Lambda \Phi_0$ and $M'_0 = h' \Lambda' \Phi'_0$,
then the Kahler potential for $\Phi_0$ and $\Phi'_0$ 
is canonical and the magnetic
quarks are canonical near the origin of field space.
Then the magnetic superpotential can be rewritten as
\bea
W_{mag} = \left[ h \Phi_0  q'  Y Y  q' 
 +  
\frac{h^2 \mu_{\phi}}{2} \tr \Phi_0^2- h \mu^2 \tr \Phi_0 \right]
+  \left[ h' \Phi'_0  q Y Y  q 
 +  
\frac{{h'}^2 \mu_{\phi}'}{2} \tr {\Phi'_0}^2- h' {\mu'}^2 \tr \Phi'_0 \right]
\nonu
\eea
where
$
\mu^2 = m \Lambda, {\mu'}^2 =m' \Lambda'$ and  
$\mu_{\phi} = \alpha \Lambda^2, \mu_{\phi}' = \alpha' {\Lambda'}^2$.

Now one splits 
the $2(N_f-l) \times 2(N_f-l)$
block  at the lower right corner of $h\Phi_0$ and $q' Y Y q'$ 
into blocks of 
size $2n$ and $2(N_f-l-n)$ and 
 one decomposes 
the $2(N_f'-l') \times 2(N_f'-l')$
block  at the lower right corner of $h' \Phi'_0$ and $q Y Y  q$ 
into blocks of 
size $2n'$ and $2(N_f'-l'-n')$
as follows \cite{GK0710}:
\bea
h\Phi_0 = \left(
\begin{array}{ccc}
0_{2l} & 0 & 0  \\
0 & h \Phi_{2n}  & 0 \\
0 & 0 & \frac{\mu^2}{\mu_{\phi}} {\bf 1}_{N_f-l-n} \otimes \sigma_3
\end{array}
\right), \qquad 
h'\Phi'_0 = \left(
\begin{array}{ccc}
0_{2l'} & 0 & 0  \\
0 & h' \Phi'_{2n'}  & 0 \\
0 & 0 & \frac{{\mu'}^2}{\mu_{\phi}'} {\bf 1}_{N_f'-l'-n'} \otimes i \sigma_2
\end{array}
\right), \nonu \\
q' Y Y  q' = \left(
\begin{array}{ccc}
\mu^2 {\bf 1}_{2l} & 0 & 0  \\
0 &  \varphi'  y y \varphi'  &  0 \\
0 & 0 & 0_{2(N_f-l-n)}
\end{array}
\right), \qquad
q  Y Y q = \left(
\begin{array}{ccc}
{\mu'}^2 {\bf 1}_{2l'} & 0 & 0  \\
0 &  \varphi y y \varphi  &  0 \\
0 & 0 & 0_{2(N_f'-l'-n')}
\end{array}
\right).
\nonu
\eea
Here $\varphi'$ is $2n \times 2(\widetilde{N}_c'-l)$
dimensional matrices and correspond to $2n$ flavors  of
the gauge group $Sp(\widetilde{N}_c'-l)$ which is unbroken and 
$\varphi$ is  $2n' \times (2\widetilde{N}_c-2l')$
dimensional matrices and correspond to $2n'$ flavors  of
the gauge group $SO(2\widetilde{N}_c-2l')$ which is unbroken.
The
$\varphi'$ corresponds to 
fundamental strings connecting the $2n$ flavor D4-branes and
$2(\widetilde{N}_c'-l)$
color D4-branes and 
$\varphi$ corresponds to 
fundamental strings connecting the $2n'$ flavor D4-branes and
$2(\widetilde{N}_c-l')$
color D4-branes.
The $\Phi_{2n}$ and $\varphi' y y \varphi'$
are $2n \times 2n$ matrices while 
$\Phi'_{2n'}$ and $ \varphi 
y y \varphi$
are $2n' \times 2n'$ matrices.
The supersymmetric ground state corresponds to
\bea
h\Phi_{2n}= \frac{\mu^2}{\mu_{\phi}} {\bf 1}_{n} \otimes \sigma_3, 
\qquad \varphi' y =0=y \varphi' \qquad \mbox{and} 
\qquad
h'\Phi'_{2n'}= \frac{{\mu'}^2}{\mu_{\phi}'} {\bf 1}_{n'} 
\otimes i \sigma_2, \qquad
y \varphi =0=\varphi y.
\nonu
\eea 

Now the full one loop potential from the above superpotential and
vacuum expectation values for the fields
takes the form
\bea
V   & = &  
|h  y \varphi' \Phi_{2n}     |^2   
  +  
| h \varphi' y y \varphi' -h \mu^2 {\bf 1}_{2n} + 
h^2 \mu_{\phi} \Phi_{2n}|^2 + b |h^2 \mu|^2 \tr \Phi_{2n} \Phi_{2n} 
\nonu \\
 & + & |h'  \Phi'_{2n'}  \varphi y    |^2   
  +  
| h' \varphi y y \varphi -h' {\mu'}^2 {\bf 1}_{2n'} + 
{h'}^2 \mu_{\phi}' \Phi'_{2n'}|^2 + b' |{h'}^2 \mu'|^2 \tr \Phi'_{2n'} \Phi'_{2n'} 
\nonu
\eea
where $b = \frac{(\ln 4-1)}{8\pi^2} \widetilde{N}_c$ and 
$b' = \frac{(\ln 4-1)}{8\pi^2} \widetilde{N}_c'$ \cite{ISS,Ahn06-1}.
Differentiating this potential with respect to 
$\Phi_{2n}$ and $\Phi'_{2n'}$ and putting $ y \varphi' =0 = \varphi y
$
\cite{GK0710}, one obtains
\bea
h \Phi_{2n} 
& \simeq & \frac{ \mu_{\phi}}{b }
{\bf 1}_n \otimes  \sigma_3 \qquad \mbox{or} \qquad
M_{2n} \simeq \frac{\alpha \Lambda^3}{\widetilde{N}_c} {\bf 1}_{n}
\otimes  \sigma_3, 
\nonu \\
h' \Phi'_{2n'} 
& \simeq & \frac{ \mu_{\phi}'}{b' }
{\bf 1}_{n'} \otimes i \sigma_2 \qquad \mbox{or} \qquad
M'_{2n'} \simeq \frac{\alpha' {\Lambda'}^3}{\widetilde{N}_c'} {\bf 1}_{n'}
\otimes i \sigma_2 
\nonu
\eea
corresponding to the $w$ coordinates of $2n$ curved flavor D4-branes between 
the $D6_{-\theta}$-branes and the NS5'-brane and  
the $w$ coordinates of $2n'$ curved flavor D4-branes between 
the $D6_{-\theta'}$-branes and the NS5'-brane respectively.

\subsection{Higher order superpotential for bifundamental}

In general, there are also different kinds of meson fields 
$M_j = Q (X X)^j Q, M'_j = Q' (X
X)^j Q'$ and $
P_r = Q (X
X)^{r} X Q' $ where $j=0, 1, \cdots,
2k'+1$ and $r=0, 1, \cdots, 2k'$ for higher order term for the
bifundamental with general rotation angles of
two $(2k'+1)$ NS5'-branes \cite{ILS}.
The magnetic superpotential contains the interaction between these
meson fields with $q, q'$ and $Y$ 
\cite{ILS}
as well as the higher order term $(Y Y)^{2k'+1}$ and mass term $Y Y$ 
which will vanish for small
$\beta, m_X$ limit. 
When the
$N_f$ $D6_{-\theta}$-branes and the extra right $2k'$ NS5'-branes 
do not create the D4-branes
and $N_f'$ $D6_{-\theta'}$-branes and the extra left $2k'$ NS5'-branes
also do not
create the D4-branes(and their mirrors),  
the extra meson fields $M_j, j\neq 0$ and $P_r$ 
do not appear in the
magnetic superpotential. 

Then 
the $k'$-dependent  reduced magnetic superpotential 
with the limit $\beta, m_X
\rightarrow 0$ 
is given by 
\bea
W_{dual}  = \left[
M_0 q' (Y Y)^{2k'+1} q' 
 +  \frac{\alpha}{2} \tr M_0^2 - m M_0  \right] 
+  \left[  M_0' q (Y Y)^{2k'+1}  q + \frac{\alpha'}{2} \tr
 {M_0'}^2 - 
m' M_0'\right].
\nonu
\eea  

Then the analysis for the previous single NS5'-brane case can be
performed in this case. 
One deforms the nonsupersymmetric brane configuration, 
which is generalized Figure 3 with an orientifold 4-plane(and 
there are multiple NS5'-branes), 
by displacing the multiple
$D6_{-\theta,-\theta'}$-branes and the left and right $NS5'$-branes 
along $v$ direction \cite{Ahn08-2,Ahn08-3}.
Then the $n$ curved flavor D4-branes attached to them are displaced
also as $k'$ different $n_j$'s connecting between 
$D6_{-\theta,j}$-brane and
$NS5_{R,j}'$-brane.
Similarly,
the $n'$ curved flavor D4-branes attached to them(as well as other
D4-branes) are displaced
also as $k'$ different $n'_{j}$'s connecting between 
$D6_{-\theta',j}$-brane and
$NS5_{L,j}'$-brane.
When we rescale the submeson fields as
$M_{j} = h \Lambda \Phi_{j} $ and $M'_{j} = h' \Lambda' \Phi'_{j}$,
then the Kahler potential for $\Phi_{j}$ and $\Phi'_{j}$ 
is canonical and the magnetic
quarks $q_j$ and $q'_{j}$  
are canonical near the origin of field space \cite{ISS}.
Then the magnetic superpotential  
can be rewritten in terms of $\Phi_j, q_j, 
\Phi'_{j}, q'_{j}, 
Y_j$ and $Y_{j'}$.
In order to see the nonsupersymmetric
meta-stable ground states, one can   
split  some of the components of $h\Phi_0, q' (Y Y)^{2k'+1} 
q', h' \Phi'_0$ and $q (Y Y)^{2k'+1}
q$.
The supersymmetric ground state corresponds to
\bea
h\Phi_{2n_j} & = & \frac{\mu_j^2}{\mu_{\phi}} {\bf 1}_{n_j} \otimes \sigma_3, 
\qquad \varphi_{n_j}' y_{n_j}^{2k'+1}  =0=   
y_{n_j}^{2k'+1}
\varphi_{n_j}' \qquad \mbox{ and} \nonu \\ 
h'\Phi'_{n'_{j}} & = & \frac{{\mu_{j}'}^2}{\mu_{\phi}'} {\bf 1}_{n_{j}'}
\otimes i \sigma_2, \qquad
y_{n_j'}^{2k'+1} \varphi_{n'_{j}} =0=\varphi_{n'_{j}} 
y_{n_j'}^{2k'+1}.
\nonu 
\eea
The full one loop potential can be written similarly
and the local nonzero stable point arises as
\bea
h \Phi_{2n_j} 
& \simeq & \frac{ \mu_{\phi}}{b_j }
{\bf 1}_{n_j} \otimes  \sigma_3 \qquad \mbox{and} \qquad
h' \Phi'_{2n_{j}'} 
 \simeq  \frac{ \mu_{\phi}'}{b_{j}' }
{\bf 1}_{n_{j}'} \otimes i \sigma_2 
\nonu
\eea
corresponding to the $w$ coordinates of $2n_j$ curved 
flavor D4-branes between 
the $D6_{-\theta,j}$-branes and the $NS5_{R,j}'$-brane and  
the $w$ coordinates of $2n_{j}'$ curved flavor D4-branes between 
the $D6_{-\theta',j}$-branes and the $NS5_{L,j}'$-brane respectively.

\section{
$SU(N_c) \times SO(N_c')$ with $N_f$-fund., $2N_f'$-vectors  
and bifund.}

Let us add an orientifold 6-plane to the brane configuration of
section 2 and one extra NS5-brane is needed for the mirror.

\subsection{Electric theory}

The type IIA supersymmetric electric
brane configuration \cite{LO,Ahn07-3} corresponding to 
${\cal N}=1$ $SU(N_c) \times SO(N_c')$ gauge theory  with  
$N_f$-fundamental flavors $Q, \widetilde{Q}$,
$2N_f'$-vectors $Q'$
and bifundamentals $X, \widetilde{X}$
can be described as two NS5-branes, two
NS5'-branes, 
$N_c$- and $N_c'$-D4-branes, $2N_f$- and $2N_f'$-D6-branes 
and $O6^{+}$-plane. 
The mass terms for the flavors 
can be achieved by displacing the D6-branes along $\pm v$
direction leading to their coordinates $v = \pm 
v_{D6_{-\theta}}( \pm v_{D6_{-\theta'}})$  
respectively while the quartic terms for the flavors 
can be obtained by rotating the D6-branes 
by an angle 
$-\theta(-\theta')$ in $(w,v)$-plane respectively. 
Then, in the electric gauge theory, the general superpotential is
given by
\bea
W_{elec} & = & 
 \frac{\alpha}{2} \tr (Q \widetilde{Q})^2 - m \tr Q \widetilde{Q}  +
 \frac{\alpha'}{2} \tr (Q' Q')^2 - m' \tr Q'
Q' \nonu \\
&+& \left[\beta_1   \tr (X \widetilde{X})^2 + \beta_2 \tr X
  \widetilde{X} \widetilde{X} X + \beta_3 (\tr X \widetilde{X})^2 
+ m_X \tr X \widetilde{X} \right]. 
\label{superelec7}
\eea
The last four terms of (\ref{superelec7}) 
are due to the rotation angles 
$\omega$ and $\omega'$ of NS5-branes and
NS5'-branes where
$\beta_i$ with $i=1, 2, 3$  depend on $\omega$ and $\omega'$ 
and the relative displacement of D4-branes where the mass term for the
bifundamental $m_X  \equiv \frac{v_{NS5'}}{2\pi \ell_s^2}$ 
is the distance along the $v$-direction.
We focus on the case   
$\beta_i(i=1, 2, 3), m_X \rightarrow 0$.

Let us summarize the ${\cal N}=1$ supersymmetric electric brane
configuration for the superpotential 
(\ref{superelec7}) 
in type IIA string theory as follows and draw this in
Figure 4:

$\bullet$
Two NS5-branes in $(012345)$ directions  

$\bullet$ 
Two NS5'-branes in  $(012389)$ directions 

$\bullet$
$N_f$ $D6_{\pm \theta}$-branes in (01237)
directions and
two other directions in $(v,w)$-plane

$\bullet$
$N_f'$ 
 $D6_{\pm \theta'}$-branes in (01237)
directions and
two other directions in $(v,w)$-plane

$\bullet$
$N_c$- and $N_c'$-color D4-branes in $(01236)$ directions   

$\bullet$ $O6^{+}$-plane in (0123789) directions

\begin{figure}[ht]
   \epsfxsize=3.0in 
\centerline{\epsffile{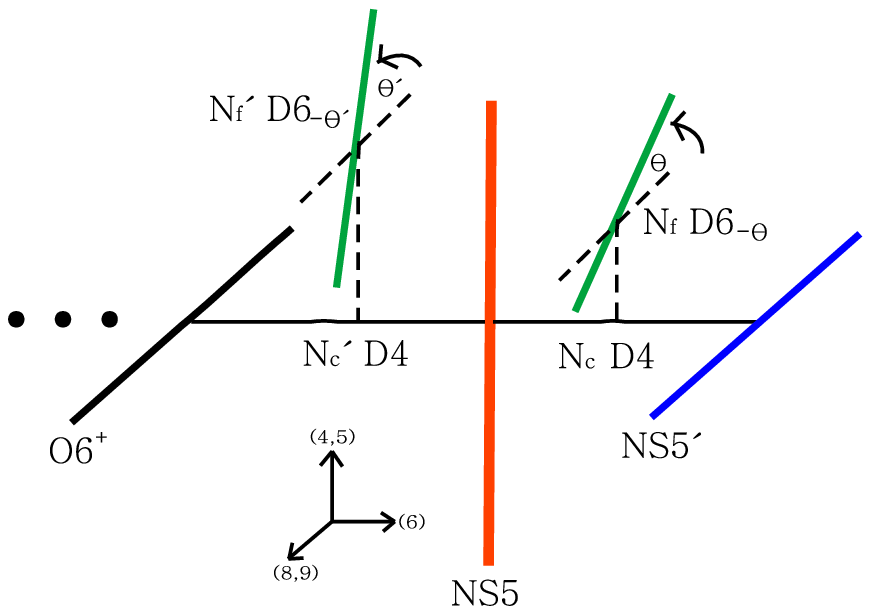}}
   \caption[FIG. \arabic{figure}.]{ 
The  ${\cal N}=1$ supersymmetric 
electric brane configuration for the gauge group $SU(N_c) \times SO(N_c')$ 
with bifundamentals $X, \widetilde{X}$ 
and fundamentals $Q, \widetilde{Q}$ and vector $Q'$. 
A 
rotation of $N_f(N_f')$ D6-branes in $(w,v)$-plane
corresponds to 
a quartic term for the fundamentals $Q, \widetilde{Q}(Q')$ while 
a displacement of $N_f(N_f')$ D6-branes in $\pm v$ direction corresponds to a
mass term for the fundamentals $Q, \widetilde{Q}(Q')$.
The mirrors located at the left hand side of O6-plane and 
denoted by $\cdots$ are preserved under the O6-plane.}
\end{figure}

\subsection{Magnetic theory}

The left NS5'-brane 
starts out with linking number $l_e=-\frac{(N_f)}{2} -\frac{4}{2} +N_c$
and after duality 
this left NS5'-brane ends up with linking number 
$l_m = \frac{(N_f)}{2} +\frac{4}{2} +N_f -\widetilde{N}_c$.
In general, when the left $N_f$ $D6_{-\theta}$-branes meet 
the left NS5-brane during the dual process,
the new D4-branes are created because they are not parallel.
However, we consider only the particular brane motion where
the left $N_f$ $D6_{-\theta}$-branes meet 
the left NS5-brane with no angles. In other words, 
the  $D6_{-\theta}$-branes become $D6_{-\frac{\pi}{2}}$-branes 
when they meet with left NS5-brane instantaneously and after that then 
they come back to the original  $D6_{-\theta}$-branes.
Therefore, in this dual process, there is no creation of D4-branes.

Also when the left $N_f$ $D6_{-\theta}$-branes meet 
the right NS5'-brane during the dual process,
the new D4-branes are created because they are not parallel.
However, we consider only the particular brane motion where
the left $N_f$ $D6_{-\theta}$-branes meet 
the right NS5'-brane with no angles.
That is, the  $D6_{-\theta}$-branes become D6-branes 
when they meet with right NS5'-brane instantaneously and then
after that  
they come back to the original  $D6_{-\theta}$-branes.
Similarly, 
when the left $N_f'$ $D6_{-\theta'}$-branes meet 
the right NS5'-brane during the dual process,
the new D4-branes are created because they are not parallel.
However, we consider only the particular brane motion where
the left $N_f'$ $D6_{-\theta'}$-branes meet 
the right NS5'-brane with no angles.
In other words, 
the  $D6_{-\theta'}$-branes become D6-branes 
when they meet with right NS5'-brane instantaneously and then 
they come back to the original  $D6_{-\theta'}$-branes.
Then it turns out that the dual color number $\widetilde{N}_c$
is given by $\widetilde{N}_c = 2N_f-N_c+4$. 

What about the other dual color number?
The left NS5-brane 
starts out with linking number $l_e=\frac{-N_f'}{2} + N_c' - N_c-\frac{4}{2}$
and after duality 
this left NS5-brane ends up with linking number 
$l_m = \frac{N_f'}{2}
+\widetilde{N}_c+N_f'-\widetilde{N}_c+\widetilde{N}_c' 
+\frac{4}{2}$.
Then it turns out that the dual color number $\widetilde{N}_c'$
is given by $\widetilde{N}_c' = 2N_f'+2N_f-N_c'+8$. 
Finally, one has the following dual color numbers 
\bea
\widetilde{N}_c = 2N_f-N_c +4, \qquad 
\widetilde{N}_c'= 2N_f'+2N_f-N_c'+8.
\nonu
\eea

The low energy theory on the color D4-branes 
has $SU(\widetilde{N}_c) \times SO(\widetilde{N}_c')$ gauge group and  
$N_f$-fundamental dual quarks $q, \widetilde{q}$
coming from 4-4 strings connecting between the color 
$\widetilde{N}_c$ D4-branes and
$N_f$ flavor D4-branes, 
$2N_f'$-vectors dual quarks $q'$
coming from 4-4 strings connecting between the color 
$\widetilde{N}_c'$ D4-branes and
$2N_f'$ flavor D4-branes
as well as $Y, \widetilde{Y}$ and various gauge singlets.
The $2N_f$ flavors $q$ and $\widetilde{q}$ and $2N_f'$ flavors $q'$ 
are in the representation 
$\bf{(\Box, 1)}, \bf{(\overline{\Box},1)}$ and $\bf{(1, \Box)}$ 
respectively under the gauge group and 
 in the representation 
$\bf{(\overline{\Box}, 1, 1)}, \bf{(1, \overline{\Box}, 1)}$
and $\bf{(1, 1, \overline{\Box})}$ 
respectively under the flavor group $SU(N_f)_L \times SU(N_f)_R
\times SU(2N_f')$.
In particular, a magnetic meson field 
\bea
M_0 \equiv Q \widetilde{Q}
\nonu
\eea
is $N_f \times N_f$ matrix and comes from 
4-4 strings of $N_f$ flavor D4-branes while
a magnetic meson field 
\bea
M'_0 \equiv Q' Q'
\nonu
\eea
is $2N_f' \times 2N_f'$ symmetric matrix and comes from 
4-4 strings of $2N_f'$ flavor D4-branes.
Then the general magnetic superpotential 
is given by 
\bea
W_{dual} & = & \left[  
 \tr (Y \widetilde{Y})^2 +  \tr Y
  \widetilde{Y} \widetilde{Y} Y + (\tr Y \widetilde{Y})^2 
+ \tr Y \widetilde{Y} \right. \nonu \\
& + & M_0 \widetilde{q} Y \widetilde{Y} Y \widetilde{Y} q + 
M_0' q' Y \widetilde{Y} \widetilde{Y} Y  q'
+ M_1 \widetilde{q} Y \widetilde{Y} q + M_1' q' Y \widetilde{Y} q' + 
M_2 \widetilde{q} q + M_2' q' q'  \nonu \\
&  + &
\left. P_0 q \widetilde{Y} Y \widetilde{Y} q' + P_1 q \widetilde{Y} q' 
 +\widetilde{P}_0 \widetilde{q} Y \widetilde{Y} Y q' +  
 \widetilde{P}_1 \widetilde{q} Y q' + R q \widetilde{Y} \widetilde{Y}
 q + \widetilde{R} \widetilde{q} Y Y \widetilde{q} \right] 
\nonu \\
& + & \frac{\alpha}{2} \tr M_0^2 - m M_0 + 
\frac{\alpha'}{2} \tr {M_0'}^2 - m' M_0'.
\label{dual}
\eea
Here the mesons 
are
$
M_1  \equiv  Q \widetilde{X} X \widetilde{Q}, 
M_1'  \equiv Q' X
\widetilde{X} Q', 
M_2   \equiv   Q \widetilde{X} X \widetilde{X} X \widetilde{Q}, 
M_2'    \equiv   Q' X \widetilde{X} \widetilde{X} X Q'$ and $ 
  P_0  \equiv  Q \widetilde{X} Q', 
 \widetilde{P}_0 \equiv \widetilde{Q} X Q', 
P_1  \equiv  Q \widetilde{X} X \widetilde{X} Q', 
\widetilde{P}_1 \equiv \widetilde{Q} X \widetilde{X} X Q', 
R  \equiv  Q \widetilde{X} \widetilde{X} Q, 
\widetilde{R} \equiv 
 \widetilde{Q} X X \widetilde{Q}$.
Note that the mesons of the $SU(N_c)$ group  couple to the dual quarks
of $SU(\widetilde{N}_c)$ and 
the mesons of the $SO(N_c')$ group  couple to the dual quarks
of $SO(\widetilde{N}_c')$, compared with the unitary case in section 2. 
The first three lines of (\ref{dual}) is present in \cite{LO}.

As in section 2, one can
dualize each gauge group independently of the other. 
One labels each NS5-branes from left to right $A,B,C$ and $D$ \cite{LO}. 
For example, one
dualizes the first gauge group factor by moving the B NS5-brane
to the left of the A NS5'-brane while the  C NS5-brane
to the right of the D NS5'-brane like as in \cite{Ahn07-3}. Then
the intermediate dual gauge group is given by 
$SU(\widetilde{n}_c =N_f+N_c'-N_c) \times SO(N_c')$ 
with corresponding superpotential for the 
dual matters. Now we interchange the A- and D- NS5'-branes and 
$D6_{-\theta,-\theta'}$-branes each other and obtain the intermediate 
dual gauge
group
 $SU(\widetilde{n}_c) \times
 SO(\widetilde{n}_c'=2N_f'+4N_f+N_c'-2N_c+4)$ 
with
dual matters. Next, we move the D NS5'-brane and
$D6_{-\theta,-\theta'}$-branes 
to the left of the B NS5-brane while
 the A NS5'-brane and
$D6_{-\theta,-\theta'}$-branes 
to the right of the C NS5-brane 
and obtain the intermediate dual
gauge group $SU(\widetilde{N}_c=4N_f+2N_f'-N_c+4) \times
SO(\widetilde{n}_c')$
with corresponding superpotential. Then, the B- and C- NS5-brane are
interchanged each other through O6-plane and we obtain the dual gauge
group
$SU(\widetilde{N}_c) \times SO(\widetilde{N}_c'=4N_f+4N_f'-N_c'+8)$
with superpotential (\ref{dual})
 \footnote{
It does not matter the order of dualization. We obtain the same
dual gauge theory if we start with $SO(N_c')$ dualization, then $SU(N_c)$
dualization, $SO(\widetilde{n}_c')$ dualization and end up with
$SU(\widetilde{n}_c)$ 
dualization.}. 

As we explained before, 
our particular brane motion during the dual process does not produce
any D4-branes 
when the left $N_f$ $D6_{-\theta}$-branes meet the left NS5-brane. This  
implies that there is no $M_2$ term in the above superpotential
(\ref{dual}).
We consider only the particular brane motion where
the left $N_f$ $D6_{-\theta}$-branes meet 
the right NS5'-brane with no angles.
This  
implies that there is no $M_1, R$ or $\widetilde{R}$ term 
in the above superpotential
(\ref{dual}). By construction, there are no additional D4-branes 
connecting the left NS5-brane and $N_f'$ $D6_{-\theta'}$-branes   
after duality. This leads to the fact that there is no $M_1'$ or $M_2'$
dependence in (\ref{dual}) \footnote{In \cite{LO}, they introduced the
$4N_f'$ full D4-branes without changing the linking number in order to
satisfy the correct dual color numbers in the gauge theory side
analysis. This has led to 
the fact that there are  $3N_f'$  D4-branes 
connecting the left NS5-brane and $N_f'$ $D6_{-\theta'}$-branes(and
their mirrors)   
after duality. In other words, the extra $2N_f'$ D4-branes were needed
for the meson fields $M_1'$ and $M_2'$. In our construction, we do not
need these extra the
$4N_f'$ full D4-branes because we do not want to have these unwanted
meson fields $M_1'$ and $M_2'$. }.
Furthermore, 
we do not see any $P_1$- or $\widetilde{P}_1$-dependent terms in the
superpotential (\ref{dual}) when the left $N_f$ $D6_{-\theta}$-branes, the 
right $N_f'$ $D6_{-\theta'}$-branes and the left NS5'-brane during the dual
process
meet each other with no angles after first $SU(N_c)$ dualization
 \footnote{These mesons 
$P_1$ and $\widetilde{P}_1$ originate from  $SU(N_c)$ chiral mesons
$Q \widetilde{X}$ and $\widetilde{Q} X$ \cite{Ahn07-3} when one
dualizes the first gauge group factor first. That is, the
strings stretching between the $N_f$ ``flavor'' D4-branes and $N_c'$
color D4-branes give rise to these $2N_f$ $SO(N_c')$ vectors. 
The superpotential in
\cite{Ahn07-3} contains the cubic term between dual bifundamental, these
meson fields and dual quarks. After three additional dual procedures, 
$SO(N_c'), SU(\widetilde{n}_c), SO(\widetilde{n}_c')$, 
these cubic terms arise as $P_1$
and $\widetilde{P}_1$-term in (\ref{dual}) where there exist  extra 
quarks $q'$-dependence while $P_1$ and $\widetilde{P}_1$ have 
extra $X \widetilde{X}Q'$ and $\widetilde{X} X Q'$ fields respectively, 
due to the further $SO(N_c')$ and $SO(\widetilde{n}_c')$-dualizations. }. 
When the left $N_f$ $D6_{-\theta}$-branes, the 
right $N_f'$ $D6_{-\theta'}$-branes and the left NS5'-brane during the dual
process
meet each other with no angles after first $SO(N_c')$ dualization, 
we do not see any $P_0$- or $\widetilde{P}_0$-dependent terms in the
superpotential (\ref{dual})
 \footnote{These mesons 
$P_0$ and $\widetilde{P}_0$ originate from  $SO(N_c')$ chiral mesons
$\widetilde{X}Q'$ and $X Q'$  when one
dualizes the second gauge group factor. That is, the
strings stretching between the $N_f'$ ``flavor'' D4-branes and $N_c$
color D4-branes give rise to these $2N_f'$ $SU(N_c)$ fundamentals. 
The superpotential contains the cubic term between dual bifundamental, these
meson fields and dual quarks. After three additional dual procedures, 
$SU(N_c), SO(\widetilde{n}_c'), 
SU(\widetilde{n}_c)$, these cubic terms arise as $P_0$
and $\widetilde{P}_0$-term in (\ref{dual}) where there exist  extra 
factors $q \widetilde{Y} Y$ and $\widetilde{q} Y \widetilde{Y}$ 
respectively while $P_0$ and $\widetilde{P}_0$ have 
extra $Q$ and $\widetilde{Q}$ fields respectively, 
due to the further $SU(N_c)$ and $SU(\widetilde{n}_c)$-dualizations. }.
Similarly, 
we do not see any $R$- or $\widetilde{R}$-dependent terms in the
superpotential (\ref{dual}) 
 \footnote{These mesons 
$R$ and $\widetilde{R}$ originate from  $SU(N_c)$ chiral mesons
$\widetilde{X}Q$ and $X\widetilde{Q}$  when one
dualizes the first gauge group factor first. That is, the
strings stretching between the $N_f$ ``flavor'' D4-branes and $N_c'$
color D4-branes give rise to these $2N_f$ $SO(N_c')$ vectors as before. 
The superpotential contains the cubic term between dual bifundamental, these
meson fields and dual quarks. After three additional dual procedures, 
$SO(N_c'), SU(\widetilde{n}_c), SO(\widetilde{n}_c')$, 
these cubic terms arise as $R$
and $\widetilde{R}$-term in (\ref{dual}) where there exist  extra 
factors $q \widetilde{Y} $ and $\widetilde{q} Y$ 
while $R$ and $\widetilde{R}$ have 
extra $Q \widetilde{X}$  and $\widetilde{Q} X$ fields, 
due to the further $SU(\widetilde{n}_c)$-dualization.}. 
Then the reduced magnetic superpotential with the limit $\beta_1,
\beta_2, \beta_3, m_X
\rightarrow 0$ 
is given by 
\bea
W_{dual}  = \left[
M_0 \widetilde{q} Y \widetilde{Y} Y \widetilde{Y} q 
 +  \frac{\alpha}{2} \tr M_0^2 - m M_0  \right] 
+  \left[  M_0' q' Y \widetilde{Y} \widetilde{Y} Y  q' 
+ \frac{\alpha'}{2} \tr
 {M_0'}^2 - 
m' M_0'\right].
\label{ddual}
\eea

For the supersymmetric vacua, one can compute the F-term equations for
this superpotential (\ref{ddual}) 
and the expectation values for $M_0, M'_0, \widetilde{q} Y
\widetilde{Y} Y \widetilde{Y}
q$ 
and $q' Y \widetilde{Y} \widetilde{Y} Y  q'$ are obtained.
The F-term equations for $M_0, q, \widetilde{q}, M_0',
q', Y$ and $\widetilde{Y}$ are 
\bea
&& \widetilde{q} Y \widetilde{Y} Y \widetilde{Y} q  -m + \alpha M_0   =  0, \qquad
(M_0 \widetilde{q} Y \widetilde{Y}) Y \widetilde{Y} =0, \qquad 
 Y \widetilde{Y} (Y \widetilde{Y} q M_0)
=0, \nonu \\
&& q' Y \widetilde{Y} \widetilde{Y} Y  q'   -m' + \alpha' M_0'  =  0, \qquad
 (M_0' q' Y \widetilde{Y}) \widetilde{Y} Y + Y \widetilde{Y}
 (\widetilde{Y} Y  q'  M_0') =0,
\nonu \\
&& Y [\widetilde{Y} \widetilde{Y} q M_0] \widetilde{q}  +  \widetilde{Y} (Y
 \widetilde{Y} q M_0) \widetilde{q}
 + q'  (M_0' q' Y 
\widetilde{Y} ) \widetilde{Y} +  \widetilde{Y} (\widetilde{Y} Y  q'
M_0') q'
  =  0, \nonu \\ 
&& q (M_0 \widetilde{q} Y \widetilde{Y}) Y + q [ M_0
 \widetilde{q} Y  Y ] \widetilde{Y} 
+ 2  Y (\widetilde{Y} Y q'  M_0' )q'
  =  0. 
\label{dualexp}
\eea
The sixth and seventh equations of (\ref{dualexp}) are satisfied if the  
the second, third, fifth  equations of (\ref{dualexp}) hold:
$
M_0 \widetilde{q} Y \widetilde{Y} = Y  \widetilde{Y} q M_0 =
M_0' q'  Y \widetilde{Y} =\widetilde{Y} Y q' M_0' =0$ 
and 
$\widetilde{Y} \widetilde{Y} q M_0 = M_0
 \widetilde{q} Y  Y=0$.
Any vacuum expectation values for the fields should obey this 
condition.
We present the magnetic brane configuration in Figure 5.

\begin{figure}[ht]
   \epsfxsize=3.5in 
\centerline{\epsffile{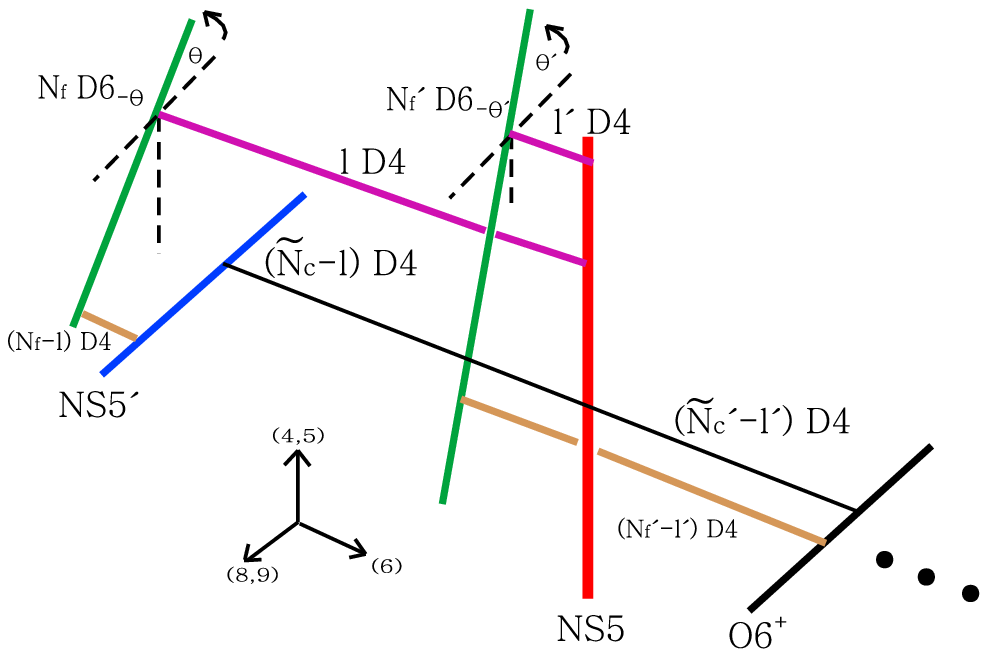}}
   \caption[FIG. \arabic{figure}.]{ 
The  ${\cal N}=1$ supersymmetric
magnetic brane configuration corresponding to Figure 4 with 
a splitting and a reconnection 
between D4-branes when the gravitational potential of
the NS5-brane is ignored. 
The $N_f$ flavor D4-branes connecting between
$D6_{-\theta}$-branes and NS5'-brane are splitting into $(N_f-l)$- and
$l$- D4-branes. 
The $N_f'$ flavor D4-branes connecting between
$D6_{-\theta'}$-branes and NS5'-brane are splitting into $(N_f'-l')$- and
$l'$- D4-branes. 
The mirrors denoted by $\cdots$ are assumed in this figure.}
\end{figure}

The theory has many nonsupersymmetric meta-stable ground states and 
when we rescale the meson field as
$
M_0 = h \Lambda \Phi_0$ and $M'_0 = h' \Lambda' \Phi'_0$,
then the Kahler potential for $\Phi_0$ and $\Phi'_0$ 
is canonical and the magnetic
quarks are canonical near the origin of field space \cite{ISS}.
Then the magnetic superpotential can be rewritten as
\bea
W_{mag} & = & \left[ h \Phi_0  \widetilde{q} Y   \widetilde{Y} Y
  \widetilde{Y} q
 +  
\frac{h^2 \mu_{\phi}}{2} \tr \Phi_0^2- h \mu^2 \tr \Phi_0 \right]
\nonu \\
&+& 
\left[ h' \Phi_0'    q' Y \widetilde{Y} \widetilde{Y} Y q' 
 +  
\frac{{h'}^2 \mu_{\phi}'}{2} \tr {\Phi_0'}^2- 
h' {\mu'}^2 \tr \Phi_0'\right]
\nonu
\eea
where
$
\mu^2 = m \Lambda, {\mu'}^2 =m' \Lambda'$ and  
$\mu_{\phi} = \alpha \Lambda^2, \mu_{\phi}' = \alpha' {\Lambda'}^2$.

Now one splits 
the $(N_f-l) \times (N_f-l)$
block  at the lower right corner of $h\Phi_0$ and $\widetilde{q} Y
\widetilde{Y} Y
\widetilde{Y} 
q$ 
into blocks of 
size $n$ and $(N_f-l-n)$ and 
 one decomposes 
the $2(N_f'-l') \times 2(N_f'-l')$
block  at the lower right corner of $h' \Phi'_0$ and $q' Y \widetilde{Y} 
\widetilde{Y} Y  q'$ 
into blocks of 
size $2n'$ and $2(N_f'-l'-n')$
as follows \cite{GK0710}:
\bea
h\Phi_0 & = &  \left(
\begin{array}{ccc}
0_l & 0 & 0  \\
0 & h \Phi_n & 0 \\
0 & 0 & \frac{\mu^2}{\mu_{\phi}} {\bf 1}_{N_f-l-n}
\end{array}
\right), \qquad
h' \Phi_0'  =   \left(
\begin{array}{ccc}
0_{2l'} & 0 & 0  \\
0 & h' \Phi_{2n'}' & 0 \\
0 & 0 & \frac{{\mu'}^2}{\mu_{\phi}'} {\bf 1}_{N_f'-l'-n'} \otimes \sigma_3
\end{array}
\right), \nonu \\
\widetilde{q}  Y \widetilde{Y} Y \widetilde{Y} q  & = & \left(
\begin{array}{ccc}
\mu^2 {\bf 1}_l & 0 & 0  \\
0 & \widetilde{\varphi} y \widetilde{y} y \widetilde{y}  \varphi  &  0 \\
0 & 0 & 0_{N_f-l-n}
\end{array}
\right), \nonu \\
 q' Y \widetilde{Y} \widetilde{Y} Y q'  & = & \left(
\begin{array}{ccc}
{\mu'}^2 {\bf 1}_{2l'} & 0 & 0  \\
0 & \varphi' y \widetilde{y} \widetilde{y} y \varphi'  &  0 \\
0 & 0 & 0_{2(N_f'-l'-n')}
\end{array}
\right).
\nonu
\eea
Here $\varphi$ and $\widetilde{\varphi}$ are $n \times (\widetilde{N}_c-l)$
dimensional matrices and correspond to $n$ flavors of fundamentals of
the gauge group $SU(\widetilde{N}_c-l)$ which is unbroken and 
$\varphi'$ is $2n' \times (\widetilde{N}_c'-l')$
dimensional matrices and correspond to $2n'$ flavors of fundamentals of
the gauge group $SO(\widetilde{N}_c'-l')$ which is unbroken 
\cite{GK0710,GK0710-1}.
The 
$\varphi$ and $\widetilde{\varphi}$ correspond to 
fundamental strings connecting the $n$ flavor D4-branes and
$(\widetilde{N}_c-l)$
color D4-branes and 
$\varphi'$ corresponds to 
fundamental strings connecting the $2n'$ flavor D4-branes and
$(\widetilde{N}_c'-l')$
color D4-branes.
The $\Phi_n$ and $\widetilde{ \varphi} y \widetilde{y} y 
\widetilde{y} \varphi$
are $n \times n$ matrices while 
$\Phi'_{2n'}$ and $\varphi' 
y \widetilde{y} \widetilde{y} y \varphi'$
are $2n' \times 2n'$ matrices.
The supersymmetric ground state corresponds to
\bea
h\Phi_n= \frac{\mu^2}{\mu_{\phi}} {\bf 1}_{n}, 
\qquad
\widetilde{\varphi} y \widetilde{y} =0=y \widetilde{y} \varphi \qquad
\mbox{ and} \qquad 
h'\Phi'_{2n'}= \frac{{\mu'}^2}{\mu_{\phi}'} {\bf 1}_{n'} \otimes \sigma_3, 
\qquad \varphi' y \widetilde{y}  =0=\widetilde{y} y \varphi'.
\nonu 
\eea 

Now the full one loop potential from the superpotential and vacuum
expectation values for the fields 
takes the form
\bea
V  & = &  
|h \Phi_n \widetilde{\varphi} y \widetilde{y} |^2   
+  |h  y \widetilde{y} \varphi \Phi_n|^2
  +  
| h \widetilde{\varphi}  y \widetilde{y} y \widetilde{y} 
\varphi-h \mu^2 {\bf 1}_{n} + 
h^2 \mu_{\phi} \Phi_n|^2 + b |h^2 \mu|^2 \tr \Phi_n^{\dagger} \Phi_n
\nonu \\ 
& + & |h' \Phi_{2n'}' \varphi'  y \widetilde{y}|^2  +
| \widetilde{y} y \varphi' \Phi'_{2n'} |^2   
  +  
|   h'  \varphi' y \widetilde{y} \widetilde{y} y 
\varphi'  -h' {\mu'}^2 {\bf 1}_{2n'} + 
{h'}^2 \mu_{\phi}' \Phi_{2n'}'|^2 \nonu \\
& + & b' |{h'}^2 \mu'|^2 
\tr \Phi'_{2n'} 
\Phi_{2n'}' 
 +  |h  \widetilde{y} \widetilde{y} \varphi \Phi_n |^2 +|h \Phi_n 
\widetilde{\varphi} y y|^2
\nonu
\eea
where $b = \frac{(\ln 4-1)}{8\pi^2} \widetilde{N}_c$ and 
$b' = \frac{(\ln 4-1)}{8\pi^2} \widetilde{N}_c'$ \cite{ISS,Ahn06-1}.
Differentiating this potential with respect to 
$\Phi_n^{\dagger}$ and ${\Phi'}_{2n'}$ and putting $\widetilde{y} 
\varphi=0=\widetilde{\varphi} y$ and $\varphi'  y=0=y \varphi'$, 
one obtains
\bea
h \Phi_n 
& \simeq &  \frac{ \mu_{\phi}}{b }
{\bf 1}_n \qquad \mbox{or} \qquad
M_n \simeq \frac{\alpha \Lambda^3}{\widetilde{N}_c} {\bf 1}_{n},
\nonu \\
h' \Phi_{2n'}' 
& \simeq &  \frac{ \mu_{\phi}'}{b' }
{\bf 1}_{n'} \otimes \sigma_3 \qquad \mbox{or} \qquad
M'_{2n'} \simeq \frac{\alpha' {\Lambda'}^3}{\widetilde{N}_c'} {\bf
  1}_{n'} \otimes \sigma_3
\nonu
\eea
corresponding to the $w$ coordinates of $n$ curved flavor D4-branes between 
the $D6_{-\theta}$-branes and the NS5'-brane and  
the $w$ coordinates of $2n'$ curved flavor D4-branes between 
the $D6_{-\theta'}$-branes and the NS5'-brane respectively.

\begin{figure}[ht]
   \epsfxsize=3.5in 
\centerline{\epsffile{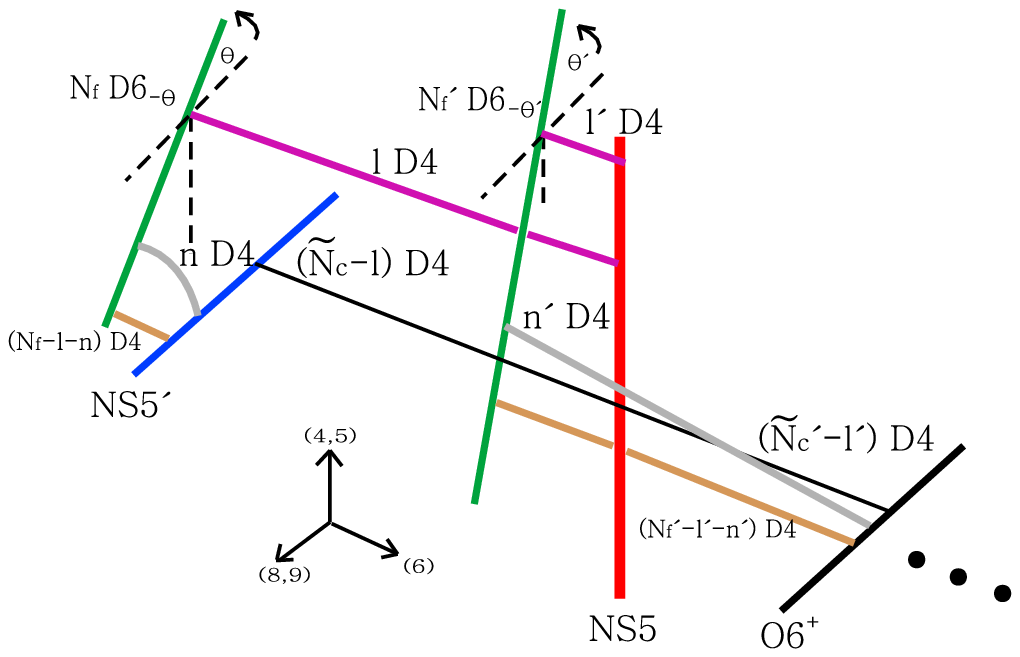}}
   \caption[FIG. \arabic{figure}.]{ 
The  nonsupersymmetric
meta-stable 
magnetic brane configuration corresponding to Figure 4 with 
a misalignment between D4-branes when the gravitational potential of
the NS5-brane is considered. 
The $(N_f-l)$ flavor D4-branes in Figure 5 connecting between
$D6_{-\theta}$-branes and NS5'-brane are splitting into $(N_f-l-n)$- and
$n$- D4-branes. 
The $(N_f'-l')$ flavor D4-branes in Figure 5 connecting between
$D6_{-\theta'}$-branes and NS5'-brane are splitting into $(N_f'-l'-n')$- and
$n'$- D4-branes. 
We do not present the mirrors denoted by $\cdots$ for simplicity.}
\end{figure}

\subsection{$SU(N_c) \times Sp(N_c')$ with 
$N_f$- and $N_f'$-fund.   
and bifund. }

In this case, the corresponding electric brane configuration can be
read off from Figure 4 by changing the $O6^{+}$-plane  into
$O6^{-}$-plane and the NS5-brane(NS5'-brane) into 
the NS5'-brane(NS5-brane).  
One analyzes the magnetic theory by using the method of the previous
case and obtains the corresponding reduced magnetic superpotential
where $M_0'$ is an antisymmetric matrix. Also the nonsupersymmetric
brane configuration can be obtained from the Figure 6 by changing the
role of NS5-brane and NS5'-brane.

\section{
Conclusions and outlook }

In this paper, we presented the meta-stable brane configurations for
the product gauge groups by dualizing the whole gauge
groups. In the context of supersymmetric brane configurations, there
are also different kinds of dual theories we did not study in this
work, from the gauge theory side or
IIA string theory side. For example, the theory with adjoint fields, 
the theory of generalization of \cite{Ahn06},
the gauge theory with triple product gauge groups which does not have
corresponding field theory analysis or higher product gauge groups
\cite{Ahn07-8,Ahn07-9}, the theory with product gauge
groups when the orientifold 6-plane is located at the NS5-brane or the
NS5'-brane \cite{Ahn07-4}, or the theory with no D6-branes 
\cite{Ahn07-5,Ahn07-6,Ahn07-7,Ahn07-10}.
It would be interesting to discover whether
the meta-stable brane configurations exist.     

\vspace{.7cm}

\centerline{\bf Acknowledgments}

This work was supported by the 
National Research Foundation of Korea(NRF) grant 
funded by the Korea government(MEST)(No. 2009-0084601).

\end{document}